\documentclass[11pt]{article}

\usepackage{amsmath}
\usepackage{amssymb}
\usepackage{times}
\usepackage{dsfont}
\usepackage{theorem}
\usepackage{graphics} 
\usepackage{cancel}
\usepackage{graphicx}

\title{Weyl, Majorana and Dirac fields from a unified perspective}
\author{Andreas Aste$^{a,b}$\\
$\quad$\\
$^{a}$\emph{Department of Physics, University of Basel, 4056 Basel, Switzerland}\\
$^{b}$\emph{Paul Scherrer Institute, 5232 Villigen PSI, Switzerland}}

\date{August 12, 2016}

\begin{document}
\maketitle

\begin{abstract}
\noindent A self-contained derivation of the formalism describing Weyl, Majorana and
Dirac fields from a unified perspective is given based on a concise description of the
representation theory of the proper orthochronous Lorentz group. Lagrangian methods play
no r\^ole in the present exposition, which covers several fundamental aspects of relativistic
field theory which are commonly not included in introductory courses when treating fermionic fields
via the Dirac equation in the first place. 
\\
\vskip 0.1 cm \noindent {\bf Physics and Astronomy Classification Scheme (2010).} 11.10.-z - Field theory;
11.30.-j Symmetry and conservation laws; 14.60.St - Nonstandard-model neutrinos, right-handed neutrinos, etc.\\

\vskip 0.1 cm \noindent {\bf Keywords.} Lorentz group; symmetry and conservation laws;
non-standard-model neutrinos; real and complex spinor representations.
\end{abstract}

\section{Introduction}
There can be great advantages in choosing a suitable notation when formulating
a theory. E.g., it is customary to denote the contravariant components of the Cartesian coordinates
of an event in flat $3+1$-dimensional Minkowski spacetime $\mathds{M} \cong (\mathds{R}^{1+3}, \eta)$ by
a column four vector
\begin{equation}
x= \left(\begin{array}{c}
x^0 \\ x^1 \\ x^2 \\ x^3
\end{array} \right) 
=
\left(\begin{array}{c}
ct \\ x^1 \\ x^2 \\ x^3
\end{array} \right) 
=
\left(\begin{array}{c}
ct \\ \vec{x}
\end{array} \right) 
\end{equation}
where the speed of light $c$ accounts for equal physical units of the
components $x^\mu$ defined by the
event time $t=x^0/c$ and the corresponding space coordinates given by a
column vector $\vec{x}=(x^1,x^2,x^3)^T$.
The Lorentz-invariant Minkowski bilinear form $\eta$ is then defined by
\begin{equation}
\eta(x,y) = x^\mu y_\mu = x^0 y^0 -x^1 y^1 - x^2 y^2 -x^3 y^3 = x^T g y
\end{equation}
with the metric tensor $g=diag(1,-1,-1,-1)$.
However, one could also hit upon the idea
to arrange the spacetime coordinates in matrix form according to \cite{Wigner}
\begin{equation}
\underline{x} = 
\left(\begin{array}{cc}
x^0+x^3 & x^1- i x^2 \\ x^1+i x^2 & x^0 -x^3
\end{array} \right) \, .
\end{equation}
Then, the indefinite Minkowski norm squared
\begin{equation}
x^2 = x_\mu x^\mu = \eta(x,x) =  (x^0)^2 - (x^1)^2 - (x^2)^2 - (x^3)^2 = \det \underline{x}
\end{equation}
can be written in an elegant manner as a determinant.
Furthermore, with
\begin{equation}
\overline{x} = 
\left(\begin{array}{cc}
x^0-x^3 & -x^1+ i x^2 \\ -x^1-i x^2 & x^0 +x^3
\end{array} \right) = 
\left(\begin{array}{cc}
x_0+x_3 & x_1- i x_2 \\ x_1+i x_2 & x_0 -x_3
\end{array} \right)
\end{equation}
one finds a compact expression for the Minkowski scalar product
\begin{equation}
\eta(x,y)=\frac{1}{2} \mbox{tr} (\underline{x} \, \overline{y}) \, .
\end{equation}
Based on notational tricks of this kind, it is possible to derive and discuss the field equations
of the fundamental fermionic fields appearing in the
Standard Model and its modern extensions in a very elegant manner, as will be demonstrated below.
Furthermore, the following discussion includes a thorough analysis of the basic properties of Weyl \cite{Weyl}, Dirac \cite{Dirac}
and Majorana \cite{Majorana}
fields emerging from first principles like Lorentz symmetry and causality.\\

In Section 2, the most relevant topological and group theoretical properties of the Lorentz group are revisited, resulting in the
construction of the fundamental ray representations of the proper orthochronous Lorentz group in Section 3.
Section 4 deals with Weyl and Majorana fields as complex two-component spinor fields. 
Section 5 contains a derivation of the Dirac equation and highlights some technical details which
are missing in the literature. Additionally, the equivalence between the complex two-component Majorana formalism
and the real four-component Majorana formalism in a Dirac setting is established in a new explicit manner.
Finally, a unified approach containing all aspects of Weyl, Majorana and Dirac fields is presented
with a special focus on the emerging Majorana phase.

\section{Structure of the Lorentz group}
In the following, Lorentz transformations will be interpreted as {\emph{passive}} transformations, i.e.
when an observer in the inertial system (or inertial frame of reference) IS assigns the contravariant Cartesian Minkowski coordinates $x^\mu$ to
an event, an observer in another inertial system IS'  with a common point of origin will assign coordinates $x'^\mu$ to the same event.
Then, the coordinates are related by a Lorentz transformation expressed by a matrix
$\Lambda$ according to $x' = \Lambda x$ or
\begin{equation}
x'^\mu = \Lambda^\mu_{\, \, \nu} x^\nu \, .
\end{equation}
Since the Minkowski metric is preserved under such transformations, one has for all $x,y \in  (\mathds{R}^{1+3}, \eta)$
\begin{equation}
\eta(x,y) = x^T g y = \eta ( \Lambda x, \Lambda y) =x^T \Lambda^T g \Lambda y \, ,
\end{equation}
and consequently
\begin{equation}
\Lambda^T g \Lambda = g \, . \label{Lo_Def}
\end{equation}
Eq. (\ref{Lo_Def}) defines the Lorentz group $\mathcal{L}$ (up to isomorphisms) as the indefinite orthogonal group $O(1,3)$
\begin{equation}
\mathcal{L} = O(1,3) = \{ \Lambda \in GL(4, \mathds{R}) \mid \Lambda^T g \Lambda = g \} \, ,
\end{equation}
where $GL(4,\mathds{R})$ denotes the multiplicative group of all invertible real $4 \times 4$-matrices. 
Since
\begin{equation}
\det(\Lambda^T g \Lambda) = \det(\Lambda^T) \det(g) \det(\Lambda) = \det(\Lambda)^2 \det(g) = \det{g} =-1
\end{equation}
one has $\det(\Lambda)=\pm 1$ for $\Lambda \in O(1,3)$. Considering matrices in $O(1,3)$
with positive determinant only, one obtains the subgroup
\begin{equation}
\mathcal{L}_+ =SO(1,3)= \{ \Lambda \in GL(4, \mathds{R}) \mid \Lambda^T g \Lambda = g, \, \det{\Lambda}=1 \}
\subset O(1,3) \, ,
\end{equation}
called the \emph{proper Lorentz group} $\mathcal{L}_+$ which is (isomorphic to) the special indefinite orthogonal group
$SO(1,3)$.\\

Representing a matrix $\Lambda \in O(1,3)$ according to the decomposition
\begin{equation}
\Lambda =
\left(\begin{array}{cc}
\gamma & -{\bf{a}}^T  \\
-{\bf{b}} & {\bf{M}} 
\end{array} \right) \, ,
\end{equation}
where $\gamma$ is a real number, ${\bf{a}}$ and ${\bf{b}}$ are column vectors, and ${\bf{M}}$ is a
$3 \times 3$ matrix, a short calculation using $(\Lambda^{-1})^T = g \Lambda g^{-1}$ gives
\begin{equation}
(\Lambda^{-1})^T =
\left(\begin{array}{cc}
\gamma & {\bf{a}}^T  \\
{\bf{b}} & {\bf{M}} 
\end{array} \right) \, .
\end{equation}
Furthermore
\begin{equation}
\gamma^2 - {\bf{b}}^2 =1 \, , \quad {\bf{b}}^T M = \gamma {\bf{a}}^T \, , \quad {\bf{M}}^T {\bf{M}} = {\bf{a}} {\bf{a}}^T+
\mathds{1}_3
\end{equation}
must hold, implying that $\gamma = \Lambda^0_{\, \, 0}$ fulfills either $\gamma \ge 1$ or $\gamma \le -1$.
Indeed, by the definition
\begin{equation}
\mathcal{L}_+^\uparrow = SO^+(1,3)= \{ \Lambda \in Mat(4, \mathds{R}) \, | \, \Lambda^T g \Lambda = g, \, \det{\Lambda}=1, \, \Lambda^0_{\, \, 0} \ge 1 \} 
\end{equation}
a further subgroup $SO^+(1,3) \subset SO(1,3) \subset O(1,3)$ is obtained.
This group, called the {\emph{proper orthochronous Lorentz group}}, is considered to be a true (local) symmetry group of all physical laws
governing quantum field theories in classical spacetimes. 
The larger group $SO(1,3)$ includes spacetime reflections (PT) with $\gamma \le -1$, and $O(1,3)$ even contains parity (P) and
time reversal (T) transformations, as discussed below. These transformations are not related to exact symmetries of the real world.
Still the group $O(1,3)$ is relevant for theoretical considerations in quantum field theory in connection with the CPT theorem \cite{Lueders}. 

\subsection{Connected components and important subgroups of the (complex) Lorentz group}
\begin{figure}[htb]
\begin{center}
\includegraphics[width=10.0cm]{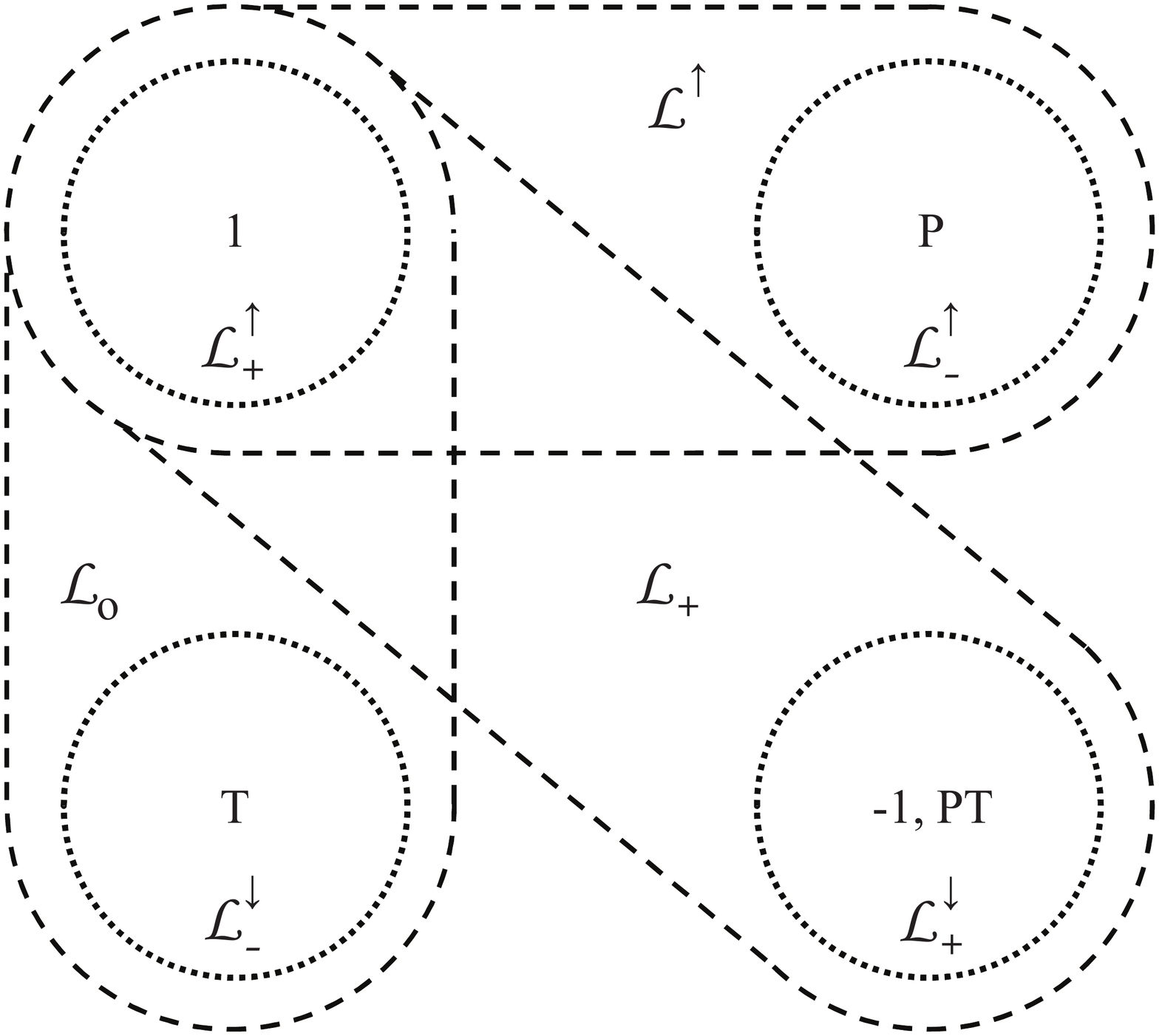}
\caption{The four pairwise disjoint and non-compact connected components of the
Lorentz group $\mathcal{L}=O(1,3)$ and corresponding subgroups:
The proper Lorentz group $\mathcal{L}_+=SO(1,3)$, the orthochronous Lorentz group
$\mathcal{L}^\uparrow$, the ortho\-cho\-rous Lorentz group $\mathcal{L}_o=\mathcal{L}_+^\uparrow
\cup T \mathcal{L}_+^\uparrow$ (see below), and the proper orthochronous Lorentz group
$\mathcal{L}_+^\uparrow=SO^+(1,3)$, which contains the identity element.
Of course, the sets $\mathcal{L}_-^\downarrow$, $\mathcal{L}_-^\uparrow$
and $\mathcal{L}_+^\downarrow$ do not represent groups due to the missing identity element.}
\label{fig_lorentz_group}
\end{center}
\end{figure}
$\mathcal{L}_+^\uparrow$ is the identity component of the Lorentz group, containing
the identity element denoted in the following simply by '$1$', by $\mathds{1}_4$, or $diag(1,1,1,1)$.
$\mathcal{L}_-^\uparrow$ contains the space reflection $P$ with $\det P = -1$,
$\mathcal{L}_-^\downarrow$ contains the time reversal  $T$ with $\det T = -1$
\begin{equation}
P =
\left(\begin{array}{cccc}
1 &0 &0 &0 \\
0 & -1 & 0 &0 \\
0 & 0 &-1 &0 \\
0 &0 & 0 & -1 \\ 
\end{array} \right) \,  , \quad 
T =
\left(\begin{array}{cccc}
-1 &0 &0 &0 \\
0 & 1 & 0 &0 \\
0 & 0 &1 &0 \\
0 &0 & 0 & 1 \\ 
\end{array} \right) \, ,
\end{equation}
$\mathcal{L}_+^\downarrow$ contains the spacetime reflection $PT= - \mathds{1}_4$ with $\det (PT) =1$.
The four transformations $\{\mathds{1}_4$, $P$, $T$, $PT \}$ constitute the discrete Klein four group $V$.\\

The so-called \emph{complex Lorentz group} $O(4,\mathds{C}) = \{ \Lambda \in
GL(4, \mathds{C}) \, | \, \Lambda^T \Lambda = \mathds{1}_4 \}$ consists of two connected components
$SO(4, \mathds{C})=O^+(4,\mathds{C})$ and $O^-(4,\mathds{C})$ only,
which can be characterized by the determinant of their elements.
The identity component $SO(4,\mathds{C})$, which contains the unimodular matrices with positive determinant, is also sometimes called 
the {\emph{proper}} complex Lo\-rentz group. One may note that the signature of the metric tensor does not play any r\^ole
for the definition of the abstract group structure of the complex Lorentz group.
A matrix $\hat{\Lambda} \in O(1,3; \mathds{C})$, which fulfills the condition
$\hat{\Lambda}^T g \hat{\Lambda}=g$ involving the metric tensor $g=diag(1,-1,-1,-1)$,
becomes via a similarity transformation
\begin{equation}
\hat{\Lambda} \rightarrow \Lambda = \Sigma \hat{\Lambda} \Sigma^{-1} \label{similar_lorentz}
\end{equation}
using
\begin{equation}
\Sigma =
\left(\begin{array}{cccc}
1 &0 &0 &0 \\
0 & i & 0 &0 \\
0 & 0 & i &0 \\
0 &0 & 0 & i \\ 
\end{array} \right)  \, , \quad
\Sigma^{-1} =
\left(\begin{array}{cccc}
1 &0 &0 &0 \\
0 & -i & 0 &0 \\
0 & 0 & -i &0 \\
0 &0 & 0 & -i \\ 
\end{array} \right)
\end{equation}
a $O(4, \mathds{C})$-matrix, since from $\Sigma^T = \Sigma$, $(\Sigma^{-1})^T = \Sigma^{-1}$
and $\Sigma^2=\Sigma^{-2}=g$ one has
\begin{equation}
\Lambda^T \Lambda = (\Sigma \hat{\Lambda} \Sigma^{-1})^T (\Sigma \hat{\Lambda} \Sigma^{-1})
= \Sigma^{-1} \hat{\Lambda}^T \Sigma \Sigma \hat{\Lambda} \Sigma^{-1}
= \Sigma^{-1} \underbrace{\hat{\Lambda}^T g \hat{\Lambda}}_{g} \Sigma^{-1} = \mathds{1}_4 \, ,
\end{equation}
and therefore $\Lambda \in O(4, \mathds{C})$.\\

A continuous path $\hat{\gamma} : [0,\pi] \rightarrow SO(1,3; \mathds{C}) \simeq SO(4,\mathds{C})$ connecting
the identity element $\mathds{1}_4$ with the spacetime reflection $PT=-\mathds{1}_4$ according to
$\hat{\gamma} (0)=\mathds{1}_4$, $\hat{\gamma} (\pi)=-\mathds{1}_4$, and 
$|\hat{\gamma}|=\{\hat{\gamma}(t) \, | \,  t \in [0,\pi]\} \subset SO(1,3;\mathds{C})$, is given by
\begin{equation}
\hat{\gamma}(t)=
\left(\begin{array}{cccc}
\cos t & i \sin t & 0 & 0 \\
i \sin t & \cos t & 0 & 0 \\
0 & 0 & \cos t & \sin t \\
0 & 0 & - \sin t & \cos t 
\end{array}\right)\; .
\end{equation}
By the similarity transformation (\ref{similar_lorentz}) the path $\hat{\gamma}$ becomes a path $\gamma$
lying completely in $SO(4, \mathds{C})$ (and even in the real group $SO(4)$)
\begin{equation}
\gamma(t) = \Sigma \hat{\gamma}(t) \Sigma^{-1}
=
\left(\begin{array}{cccc}
\cos t &  \sin t & 0 & 0 \\
- \sin t & \cos t & 0 & 0 \\
0 & 0 & \cos t & \sin t \\
0 & 0 & - \sin t & \cos t 
\end{array}\right)
\, .
\end{equation}

Of course, also the pseudo-unitary group $U(1,3) \not\simeq U(4)$ could be considered as a complex
generalization of the proper Lorentz group; the corresponding group manifold would be 16-dimensional,
whereas the topological dimension of the group manifold $SO(4, \mathds{C})$ is 12.
$U(1,3)$ is the symmetry group of a 'Minkowski sesquilinear form',
but it turned out that the group  $SO(4, \mathds{C})$ leads to more fruitful applications in quantum field theory
in connection with the analytic continuation of certain theoretical constructs like, e.g., Wightman correlation distributions
\cite{PCT}.

\section{Fundamental ray representations of the proper orthochronous Lorentz group}
\subsection{Explicit construction of the two two-dimensional inequivalent irreducible fundamental ray representations}
In order to construct the lowest-dimensional non-trivial irreducible representations of the
proper orthochronous Lorentz group, one may introduce the relativistically generalized Pauli matrices
\begin{equation}
\sigma_\mu = \bar{\sigma}^\mu=(\mathds{1}_2,\vec{\sigma})=(\sigma_0, \sigma_1, \sigma_2, \sigma_3) \label{sigma4} \, , \quad
\bar{\sigma}_\mu=\sigma^\mu=(\mathds{1}_2 ,-\vec{\sigma})
\end{equation}
where the three components in $\vec{\sigma}$ are given by the Pauli matrices
\begin{equation}
\sigma_1=\left(\begin{array}{rr} 0 & 1 \\
1 & 0 \end{array}\right) \; , \quad
\sigma_2=\left(\begin{array}{rr} 0 & -i \\
i & 0 \end{array}\right) \; , \quad
\sigma_3=\left(\begin{array}{rr} 1 & 0 \\
0 & -1 \end{array}\right) \, .
\end{equation}
The symbol $\sigma_0$ used for the identity matrix $\mathds{1}_2$ in two dimensions
fits nicely into the notation used above.\\
\vskip -0.2cm
Arbitrary four vectors $x$ with contravariant components
$x^\mu=(x^0,x^1,x^2,x^3)$ are mapped by the linear bijection $x \mapsto \underline{x}$
\begin{equation}
Herm(2, \mathds{C}) \ni \underline{x}=\sigma_\mu x^\mu=
\left(\begin{array}{cc} x^0+x^3 & x^1-ix^2 \\
x^1+ix^2 & x^0-x^3 \end{array}\right) \;  \label{linbij}
\end{equation}
onto the set $Herm(2, \mathds{C})$ of Hermitian $2 \! \times \! 2$ matrices.
The inverse mapping can be easily obtained from a generalization of the well-know trace identity
\begin{equation}
\frac{1}{2} \mbox{tr}\,  (\sigma_i \sigma_j) = \delta_{ij} \, ,
\end{equation}
i.e.
\begin{equation}
\frac{1}{2} \mbox{tr}\, (\sigma_\mu \bar{\sigma}_\nu) = g_{\mu \nu} \, .
\end{equation}
Then one has
\begin{equation}
\frac{1}{2} \mbox{tr}\, (\underline{x} \bar{\sigma}^\mu) = \frac{1}{2} \mbox{tr}\, ( x^\nu \sigma_\nu \bar{\sigma}^\mu ) = x^\nu g_\nu^{\, \, \mu}
= x^\mu
\end{equation}\\
The Minkowski scalar product can be transferred from $\mathds{M}$ onto $Herm(2 , \mathds{C})$
via
\begin{equation}
\mbox{det} \, \underline{x}=(x^0+x^3)(x^0-x^3)-(x^1+ix^2)(x^1-ix^2)=(x^0)^2-(x^1)^2-(x^2)^2-(x^3)^2=x_\mu x^\mu = x^2 
\end{equation}
and the identity
\begin{equation}
\eta(x+y,x+y)=(x+y,x+y)_M= (x+y)_\mu (x+y)^\mu = x^2 + y^2 +2 x_\mu y^\mu \, ,
\end{equation}
leading to
\begin{equation}
\eta(x,y) = x_\mu y^\mu = \frac{1}{2} ( \det ( \underline{x}+\underline{y})-\det (\underline{x})-\det (\underline{y})) \, .
\end{equation}
Alike, with $\overline{y} = \bar{\sigma}_\mu y^\mu$ one obtains the compact expression for the Minkowski scalar product
\begin{equation}
\eta(x,y) = \frac{1}{2} \mbox{tr}\,  (\underline{x} \, \overline{y}) \, .
\end{equation}
The \emph{special linear group} $SL(2,\mathds{C})$ in two complex dimensions is defined by
\begin{equation}
SL(2,\mathds{C})=\{A \! \in \! GL(2,\mathds{C}) \, | \,
\mbox{det} A = +1 \}.
\end{equation}
Now, a handy trick relies on the possibility to act with a matrix $A \in SL(2,\mathds{C})$ on $\underline{x} \in Herm(2, \mathds{C})$ according to
\begin{equation}
\underline{x} \mapsto \underline{x}'=A \underline{x} A^+ \label{trick}
\end{equation}
where $^+$ denotes Hermitian conjugation. Obviously,
$\underline{x}'$ is Hermitian again, and the Minkowski scalar product is preserved in the following sense
\begin{equation}
\mbox{det} \, \underline{x}' = \mbox{det} (A \underline{x} A^+)= 
\mbox{det} \, A \, \mbox{det} \, \underline{x} \, \mbox{det} \, A^+ = \mbox{det} \, \underline{x} \, .
\end{equation}
Thus again, $\underline{x}'$ can be represented by a real linear combination of generalized Pauli matrices
\begin{equation}
\underline{x}'=\sigma_\mu x'^\mu \, \,  \, \, \mbox{with}
\, \, x'_\mu x'^\mu=x_\mu x^\mu
\end{equation}
and $A$ explicitly acts as a Lorentz transformation due to
\begin{equation}
x'^\mu = \frac{1}{2} \,  \mbox{tr}\, (\underline{x}' \bar{\sigma}^\mu) = \frac{1}{2}  \, \mbox{tr}\, 
(A x^\nu \sigma_\nu A^+ \bar{\sigma}^\mu )
=\frac{1}{2} \, \mbox{tr}\, (A \sigma_\nu A^+ \bar{\sigma}^\mu ) x^\nu = \Lambda^\mu_{\, \, \nu} x^\nu \, .
\label{so13_sl2c}
\end{equation}
Since $x$ and $x'$ are obviously related by a Lorentz transformation, one has $\Lambda \in O(1,3)$,
and a closer inspection shows that $\Lambda \in SO^+(1,3)$ holds indeed.
Actually,  $SL(2,\mathds{C})$ is simply connected as will be demonstrated later. Since the mapping $\lambda: A \mapsto
\Lambda(A)$ is obviously continuous, it is also a homomorphism
of the group $SL(2,\mathds{C})$ into the proper orthochronous Lorentz group
$\mathcal{L}_+^{\uparrow}=SO^+(1,3)$. Furthermore, $\lambda$ is surjective, and
$SL(2, \mathds{C})$ is the double universal covering group of the $SO^+(1,3)$.\\

Mapping covariant components $x_\mu$  of a four vector according to
$\overline{x} = \overline{\sigma}^\mu x_\mu =\overline{\sigma}_\mu x^\mu$
onto $Herm(2,\mathds{C})$, the transformation law corresponding to eq. (\ref{trick}) reads
$\overline{x}'=(A^+)^{-1} \overline{x} A^{-1}$. Then one also has
\begin{equation}
x'^\mu y'_\mu= \frac{1}{2} \mbox{tr} (A \underline{x} A^+ (A^+)^{-1} \overline{y} A^{-1}) 
= \frac{1}{2} \mbox{tr} (A^{-1} A \underline{x} \, \overline{y}) =  \frac{1}{2} \mbox{tr}
\underline{x} \, \overline{y} = x^\mu y_\mu.
\end{equation}

To convince oneself that the homomorphism $\lambda$ is two-to-one, one observes first
that two matrices $\pm A \in SL(2, \mathds{C})$ generate the same Lorentz transformation,
since $A \underline{x} A^+ = (-A) \underline{x} (-A)^+$.
The kernel of $\lambda$, i.e., the set of all $A \! \in \! SL(2,\mathds{C})$ which
fulfill the equation
\begin{equation}
\underline{x}=A \underline{x}  A^+ \label{condition}
\end{equation}
for every Hermitian matrix $\underline{x}$, can be determined by first considering the
special choice
\begin{displaymath}
\underline{x}=\left(\begin{array}{rr} 1 & 0 \\
0 & 1 \end{array}\right) \; ,
\end{displaymath}
which leads to the condition $A=(A^+)^{-1}$ for $A \in ker(\lambda)$, and
eq. (\ref{condition}) then reduces to $\underline{x} A-A\underline{x}=[ \underline{x}  ,  A ]=0$
for every Hermitian $\underline{x}$.
This implies $A=\alpha \mathds{1}_2$, $\alpha \in \mathds{C}$. Eventually, from the condition
$\mbox{det} \, A=+1$ follows $A=\pm \mathds{1}_2$.\\
\vskip -0.2cm

Aside from the important group isomorphism just found above
\begin{equation}
\mathcal{L}^\uparrow_+ \cong SL(2,\mathds{C})/\{\pm 1\} \, , \quad
\underline{\Lambda(\pm A)x} = A \underline{x} A^+ \, ,
\end{equation}
the matrices in $SL(2,\mathds{C})$ display a further interesting property.
Defining the fully anti-symmetric tensor $\epsilon=-\epsilon^{-1}=
-\epsilon^{\mbox{\tiny{T}}}$ in two dimensions by
\begin{equation}
\epsilon=i \sigma_2=\left(\begin{array}{rr} 0 & 1 \\
-1 & 0 \end{array}\right) \; ,
\end{equation}
a \emph{symplectic} (and therefore \emph{skew-symmetric}) bilinear form
$\langle u ,v \rangle =-\langle v , u \rangle$ can be defined on two so-called
{\emph{spinors}} $u$ and $v$, which are elements of the two-dimensional complex vector
(or spinor) space $\mathds{C}^2_{_{\mathds{C}}}$
\begin{equation}
u=\left(\begin{array}{cc} u^1 \\ u^2 \end{array}\right) \; , \quad
v=\left(\begin{array}{cc} v^1 \\ v^2 \end{array}\right)
\end{equation}
equipped with the symplectic form $\langle \cdot , \cdot \rangle$ according to
\begin{equation}
\langle u , v \rangle = u^1 v^2 - u^2 v^1=
u^{\mbox{\tiny{T}}} \epsilon v \, .
\end{equation}
In analogy to the $SO^+(1,3)$-invariance of the metric tensor $g$,
this symplectic form is $SL(2,\mathds{C})$-invariant
\begin{equation}
\langle u , v \rangle = u^{\mbox{\tiny{T}}} \epsilon v=
\langle Au , Av \rangle = u^{\mbox{\tiny{T}}} A^{\mbox{\tiny{T}}}
\epsilon A v \, .
\end{equation}
This can be easily demonstrated by a short calculation. With
\begin{equation}
A=\left(\begin{array}{rr} a^1_{\, \, 1} & a^1_{\, \, 2} \\
a^2_{ \, \, 1} & a^2_{ \, \, 2} \end{array}\right) \; , \quad
\mbox{where} \quad
\mbox{det} \, A=a^1_{\, \, 1} a^2_{ \, \, 2}-a^1_{\, \, 2}
a^2_{ \, \, 1} \, ,
\end{equation}
one has
\begin{displaymath}
A^{\mbox{\tiny{T}}} \epsilon A=
\left(\begin{array}{rr} a^1_{\, \, 1} & a^2_{\, \, 1} \\
a^1_{ \, \, 2} & a^2_{ \, \, 2} \end{array}\right)
\left(\begin{array}{rr} 0 & 1 \\
-1 & 0 \end{array}\right) 
\left(\begin{array}{rr} a^1_{\, \, 1} & a^1_{\, \, 2} \\
a^2_{ \, \, 1} & a^2_{ \, \, 2} \end{array}\right)=
\end{displaymath}
\begin{equation}
\left(\begin{array}{rr} a^1_{\, \, 1} & a^2_{\, \, 1} \\
a^1_{ \, \, 2} & a^2_{ \, \, 2} \end{array}\right)
\left(\begin{array}{cc} a^2_{\, \, 1} & a^2_{\, \, 2} \\
-a^1_{ \, \, 1} & -a^1_{ \, \, 2} \end{array}\right)=
\left(\begin{array}{cc} 0 & \mbox{det} \, A \\
-\mbox{det} \, A & 0 \end{array}\right)=
\left(\begin{array}{rr} 0 & 1 \\
-1 & 0 \end{array}\right) \; ,
\end{equation}
such that a further group isomorphism is established
\begin{equation}
\mathcal{L}^\uparrow_+ \cong Sp(2,\mathds{C})/\{ \pm 1 \}, \quad
SL(2,\mathds{C}) \cong Sp(2,\mathds{C}) \, .
\end{equation}
$Sp(2,\mathds{C})$ is the complex symplectic
group in two dimensions
\begin{equation}
Sp(2,\mathds{C})=\{ A \! \in \! GL(2,\mathds{C}) \, | \,
A^{\mbox{\tiny{T}}} \epsilon A=\epsilon \} \, .
\end{equation}
The accidental isomorphism
$SO(3,\mathds{C}) \cong \mathcal{L}^\uparrow_+$
is mentioned here without proof for the sake of completeness.\\
\vskip -0.2cm

To sum up, the trick expressed by eq. (\ref{trick}) serves to construct
a real linear representation of the Lie group $SL(2,\mathds{C})$
by the Lie group $SO^+(1,3)$ with the defining property for representations
\begin{equation}
\Lambda(A_1) \Lambda(A_2) = \Lambda(A_1 A_2) \, . \label{proper_rep}
\end{equation}
Eq. (\ref{proper_rep}) can be inverted up to a sign, and in a loose style one may write
$A(\Lambda_1 \Lambda_2)=\pm A(\Lambda_1) A(\Lambda_2)$, i.e. also a two-valued
\emph{ray representation} of the Lorentz group $\mathcal{L}_+^\uparrow$ has been found.\\

The inversion of eq. (\ref{so13_sl2c}), which fails for some $\Lambda$ for topological reasons,
reads
\begin{equation}
A=\pm \frac{\Lambda^\mu_{\, \, \nu} \sigma_\mu \bar{\sigma}^\nu}
{\sqrt{\det  (\Lambda^\mu_{\, \, \nu} \sigma_\mu \bar{\sigma}^\nu})} \, . \label{inv_rep}
\end{equation}
The derivation of eq. (\ref{inv_rep}) is left to the reader as an exercise.\\

One may ask whether the two-valued ray representation
of the proper orthochronous Lorentz group or the representation
of the $SL(2, \mathds{C})$ by itself is equivalent to the complex conjugate representation.
It turns out that no matrix $B$ exists such that
for all $A \! \in \! SL(2,\mathds{C})$
\begin{equation}
A^*=B A B^{-1}
\end{equation}
holds. However, restricting our considerations to the $SL(2, \mathds{C})$-subgroup
\begin{equation}
SU(2)=\{ U \! \in \! GL(2,\mathds{C}) \, | \, U^+=U^{-1}, \,
\mbox{det} \, U=1 \} \! \subset \! SL(2,\mathds{C}) \, ,
\end{equation}
the situation is different, because a special unitary matrix $U \! \in \! SU(2)$ given by
\begin{equation}
U=\left(\begin{array}{cc} a & b \\
-b^* & a^* \end{array}\right) = (U^{-1})^+ \; , \quad
\mbox{det} \, U=aa^*+bb^*=1 
\end{equation}
is related to its complex conjugate matrix $U^*$ by a similarity transformation
expressed by the help of the anti-symmetric tensor $\epsilon= i \sigma_2$
\begin{equation}
\epsilon U \epsilon^{-1} =
\left(\begin{array}{rr} 0 & 1 \\
-1  & 0 \end{array}\right) 
\left(\begin{array}{cc} a & b \\
-b^* & a^* \end{array}\right)
\left(\begin{array}{rr} 0 & -1 \\
1  & 0 \end{array}\right) 
=
\left(\begin{array}{rr} a^* & b^* \\
-b  & a \end{array}\right) 
=U^* \, .
\end{equation}

Therewith the ray representation of the rotation group $SO(3)$ by the special unitary group
$SU(2)$, obtained from the restriction of the $SL(2,\mathds{C})$ representation constructed above
via the trick in eq. (\ref{trick})
to the subgroup $SU(2)$, turns out to be equivalent to its complex conjugate representation.
The $SU(2)$ matrices in the $SL(2, \mathds{C})$ generate spatial rotations;
Hermitian matrices in the $SL(2,\mathds{C})$ generate the boosts.

We will denote the fundamental representation of the spinor Lorentz group $SL(2, \mathds{C})$ by itself
as the $(\frac{1}{2},0)$-representation and its complex conjugate representation $(\frac{1}{2},0)^*$ as the
$(0,\frac{1}{2})$-representation. The trivial irreducible representation is the $(0,0)$-representation.

\subsection{Wigner boosts}
A Wigner boost is a proper orthochronous Lorentz transformation,
which transforms a given four vector $q$ into another four vector $p$.
For illustrative purposes only, the special case with $m>0$
\begin{equation}
q=\left(\begin{array}{c} 
m \\
0\\
0\\
0\\
\end{array}\right)  
\overset{Wigner \, \, boost}{\longrightarrow} 
p=\left(\begin{array}{c} 
p^0 \\
p^1\\
p^2\\
p^3\\
\end{array}\right)  
\end{equation}
shall be investigated here.
Since $q$ is a future-directed, timelike vector, $p$ is also contained in the open forward light-cone
and one has $p^2 =p_0^2-|\vec{p} \, |^2 =m^2 >0$ and $p^0=p_0 \ge m >0$. \\

With $\underline{q}= \sigma_\mu q^\mu = m \mathds{1}_2$ and
\begin{displaymath}
A_{_{W}} = \sqrt{\underline{p}/m} = \frac{1}{\sqrt{m}} \sqrt{p^0 \mathds{1}_2 + \vec{\sigma} \cdot \vec{p}}
\end{displaymath}
\begin{equation}
= \pm \frac{1}{2 \sqrt{m}} \Bigl[ \bigl( \sqrt{p^0 +|\vec{p} \, |}+ \sqrt{p^0 - |\vec{p} \, |} \bigl) \mathds{1}_2 
+ \bigl( \sqrt{p^0 + |\vec{p} \, |} - \sqrt{p^0 - |\vec{p} \, |}  \bigl) \vec{\sigma} \cdot \frac{\vec{p}}{| \vec{p} \, |}
\Bigl] \, , \label{squareroot_wigner}
\end{equation}
a Wigner boost is given, since
\begin{equation}
A_{_{W}} \underline{q} A_{_{W}}^+ = m A_{_W} A_{_W}^+ =m \sqrt{\underline{p}/m}^2 = \underline{p} \, .
\end{equation}
The explicit expression for the square root in eq. (\ref{squareroot_wigner}) can be easily verified by a short
calculation. For $m=0$ the expression becomes useless, signaling fundamental differences between the
physics of massive and massless particles.

\subsection{Topology of the $SL(2,\mathds{C})$ group manifold}
An invertible matrix $A \in GL(n, \mathds{C})$ possesses the polar decomposition
\begin{equation}
 A = H \cdot U \, ,
\end{equation}
where $H=H^+ \in Herm(n, \mathds{C})$ is a {\emph{positive definite}} Hermitian matrix and
$U=(U^{-1})^+ \in U(n)$ is unitary.
For $A \in GL(1, \mathds{C}) = \dot{\mathds{C}}$, the polar decomposition reduces to the well-known form
\begin{equation}
z=H \cdot e^{i \varphi} \, , \, \, 0 < H \in \mathds{R}^+ \, , \, \,  e^{i \varphi} \in U(1) \, .
\end{equation}
Singular matrices can be represented by the product of a unitary and a positive semi-definite Hermitian matrix.

The existence of the polar decomposition for matrices follows from the observation that the matrix
 $\tilde{H}=A A^+$ is Hermitian, since $(A A^+)^+=A^{++} A^+ = A A^+$. Obviously, $\tilde{H}$ is also invertible
when $A$ is invertible.
If $v$ is an eigenvector of $A A^+$ with a corresponding eigenvalue
$\lambda$, then due to Hermiticity $\lambda$ is real and from
\begin{equation}
\lambda v^+ v = v^+ A A^+ v = (A^+ v)^+ (A^+ v) \ge 0
\end{equation}
one even has $\lambda > 0$. Since $\det A \neq 0$, $\lambda_{1, \ldots n} > 0$ holds for {\emph{all}} $n$
eigenvalues $\lambda_1 , \ldots \lambda_n$ of $A A^+$.\\
Diagonalizing $A A^+$ by a unitary matrix $\tilde{U} \in U(n)$ leads to the real diagonal matrix
\begin{equation}
D= \tilde{U} \tilde{H} \tilde{U}^{-1} =
\left(\begin{array}{cccc}
\lambda_1 &  & & \\
  & \ldots & & \underline{0} \\
\underline{0} & & \ldots & \\
 & & & \lambda_n
\end{array} \right)
 \, ,
\end{equation}
accordingly $\tilde{H} = \tilde{U}^{-1} D \tilde{U}$ also holds.
Now, choosing $H$ and $U$ as follows
\begin{equation}
H = \tilde{U}^{-1} \sqrt{D} \tilde{U} \, , \quad U = H^{-1} A \, , \quad
\sqrt{D}=
\left(\begin{array}{cccc}
\sqrt{\lambda_1} &  & & \\
  & \ldots & & \underline{0} \\
\underline{0} & & \ldots & \\
 & & & \sqrt{\lambda_n} 
\end{array} \right)
\end{equation}
directly leads to the desired polar decomposition, since
$H^+ = \tilde{U}^+ \sqrt{D} (\tilde{U}^{-1})^+ = H$
is Hermitian and $U$ is unitary because of
\begin{equation}
U U^+ = H^{-1} A A^+ (H^{-1})^+ =
\underbrace{\tilde{U}^{-1} \sqrt{D^{-1}} \tilde{U}}_{H^{-1}} \cdot
\underbrace{\tilde{U}^{-1} D \tilde{U}}_{A A^+ = \tilde{H}} \cdot
\underbrace{\tilde{U}^{-1} \sqrt{D^{-1}} \tilde{U}}_{(H^{-1})^+} = \mathds{1}_n \, .
\end{equation}

For the special case $A \in SL(2, \mathds{C}) \subset GL(2, \mathds{C})$ follows a unique
decomposition with $\det \, H = 1$ and $U \in SU(2)$, since
\begin{equation}
\det A =1 = \underbrace{\det H}_{=\lambda_1 \cdot  \lambda_2 \in \mathds{R}^+} \cdot
\underbrace{\det U}_{|\det{U}|=1} \quad \rightarrow \quad \det{H}=1 \, , \quad U \in SU(2) \, .
\end{equation}
The representation
\begin{equation}
H = h^\mu \sigma_\mu \, , \quad \det{H}=h_\mu h^\mu = h_0^2-\vec{h}^2 =1 \, ,
\end{equation}
where $\vec{h} \in \mathds{R}^3$ can be chosen in an arbitrary manner, implies $h_0^2 = 1 + \vec{h}^2 \ge 0$.
Furthermore, $h^0$ is fully determined by $\vec{h}$ since $2 h^0 =\mbox{tr} \, H = \lambda_1 + \lambda_2 \ge 0$,
i.e.
\begin{equation}
h^0 = \sqrt{1 + \vec{h}^2} \, ,
\end{equation}
implying the homeomorphism
\begin{equation}
SL(2, \mathds{C}) \cong \mathds{R}^3 \times S^3 \, , \label{topprod}
\end{equation}
since $SU(2) \cong S^3$. Since both manifolds $\mathds{R}^3$ and $S^3$ are simply connected,
the same observation follows for the group manifold $SL(2, \mathds{C})$ as a topological product space.
Whereas the group manifold of the $SU(2)$ is homeomorphic to the compact three-dimensional sphere $S^3$,
the Lorentz group manifold is non-compact due to the non-compact factor $\mathds{R}^3$ in eq. (\ref{topprod}).
Is is assumed here that the reader
is well acquainted with the basic topological facts concerning the manifolds and matrix Lie groups equipped with their
standard topologies discussed so far.\\

All four components of the Lorentz group $O(1,3)$ like the $SO^+(1,3)$ are not simply connected,
but the covering group $SL(2, \mathds{C})$ of $SO^+(1,3)$ is. This topological difference and the related
two-to-one surjective homomorphism of the $SL(2, \mathds{C})$ onto the proper orthochronous Lorentz group
discussed above is the origin of spinor physics,
a fact which is deeply related to Wigner's theorem where physical states are related to rays in a Hilbert space \cite{Bargmann}.
However, we will not dwell any further on the specific aspects of Wigner's theorem in this paper.

\section{Spin-$\frac{1}{2}$: Two-component spinor wave equations}
\subsection{Weyl equations}
Remembering the Wigner trick (\ref{linbij}) relying on the bijection
\begin{equation}
\mathds{M} \ni x \mapsto \underline{x}=\sigma_\mu x^\mu \in Herm(2, \mathds{C})
\end{equation}
and the transformation (\ref{trick})
\begin{equation}
\underline{x}' = (\pm A) \underline{x} (\pm A^+) = A \sigma_\mu x^\mu A^+ =
A \sigma_\mu A^+  x^\mu = \sigma_\mu x'^\mu = \sigma_\mu \Lambda^\mu_{\, \, \nu}
x^\nu = \sigma_\nu \Lambda^\nu_{\, \, \mu} x^\mu \label{pauli_L_trafo} \,  ,
\end{equation}
we now use the fact that a two-to-one correspondence $\Lambda (\pm A) \leftrightarrow \pm A(\Lambda)$
exists between proper orthochronous Lorentz transformations $\Lambda \in SO^+(1,3) = \mathcal{L}_+^\uparrow$
and two corresponding elements $\pm A$ of the special linear group $SL(2, \mathds{C})$ in order to construct
the most fundamental spinor wave equations.\\

Obviously, eq. (\ref{pauli_L_trafo}) implies
\begin{equation}
A \sigma_\mu A^+ = \sigma_\nu \Lambda^\nu_{\, \, \mu} \, . \label{sigma_L_pauli}
\end{equation}
In analogy to the construction given by eq. (\ref{linbij}), one defines the matrix-valued differential operator
\cite{Aste1}
\begin{equation}
\underline{\partial} = \sigma_\mu \partial^\mu =
\left(\begin{array}{cc}
\partial^0 + \partial^3 & \partial^1 - i \partial^2  \\
\partial^1 + i \partial^2 & \partial^0 - \partial^3   \\
\end{array} \right)  \, , \quad i^2=-1 \, .  \label{weylop1}
\end{equation}
E.g., this operator can act by a formal matrix multiplication from the left on a two-component wave function
\begin{equation}
\psi(x) =
\left(\begin{array}{c}
\psi_1 (x)  \\
\psi_2 (x)   \\
\end{array} \right) \in \mathds{C}^2 \, ,
\end{equation}
explicitly
\begin{equation}
\left(\begin{array}{cc}
\partial^0 + \partial^3 & \partial^1 - i \partial^2  \\
\partial^1 + i \partial^2 & \partial^0 - \partial^3   \\
\end{array} \right) 
\left(\begin{array}{c}
\psi_1  \\
\psi_2   \\
\end{array} \right) =
\left(\begin{array}{c}
\partial^0 \psi_1 + \partial^3 \psi_1 + \partial^1 \psi_2 - i \partial^2 \psi_2 \\
\partial^0 \psi_2 - \partial^3 \psi_2 + \partial^1 \psi_1 + i \partial^2 \psi_1   \\
\end{array} \right) \, .
\end{equation}
Ignoring the trivial case where the two components of the wave function individually behave as scalar
wave functions
\begin{equation}
\psi'(x') = \left(\begin{array}{c}
\psi'_1 (x')  \\
\psi'_2 (x')   \\
\end{array} \right) 
= \psi(x) =
\left(\begin{array}{c}
\psi_1 (x)  \\
\psi_2 (x)   \\
\end{array} \right) \, ,
\end{equation}
the spinor components of the wave function $\psi(x)$ passively transform equivalently to a irreducible fundamental
ray transformation $(\frac{1}{2},0)$ or $(0,\frac{1}{2})$ in the following sense
\begin{equation}
(\frac{1}{2},0) : \, \psi (x) \rightarrow  \psi' (x') = S A(\Lambda) S^{-1} \psi (x) =
 S A(\Lambda) S^{-1} \psi (\Lambda^{-1} x') \label{righttrafo}
\end{equation}
or
\begin{equation}
(0,\frac{1}{2}) = (\frac{1}{2},0)^* : \, \psi (x) \rightarrow  \psi' (x') =
S' A^* (\Lambda) S'^{-1} \psi (x) \, ,
\label{lefttrafo}
\end{equation}
with fixed invertible matrices $S, S' \in GL(2, \mathds{C})$, such that one has of course
$S A S^{-1} \in SL(2, \mathds{C}) \ni S' A^* S'^{-1}$, since $\det (S A S^{-1}) = \det A = \det (S' A^* S'^{-1})$.\\

An observer in an inertial system IS' using coordinates $x'^\mu$ will use the operator
(\ref{weylop1}) according $\underline{\partial}' = \sigma_\mu \partial'^\mu$.
With eq. (\ref{sigma_L_pauli}), one immediately notes the corresponding transformation law
\begin{equation}
\underline{\partial'} = \sigma_\mu \partial'^\mu =  \sigma_\mu \Lambda^\mu_{\, \, \nu} \partial^\nu =
\sigma_\nu \Lambda^\nu_{\, \, \mu} \partial^\mu = A \sigma_\mu A^+ \partial^\mu = A \sigma_\mu \partial^\mu A^+=
A \underline{\partial} A^+ \label{traflaw1} \, .
\end{equation}

The following simple differential equation for a two-component, necessarily complex spinor wave function $\psi(x)$
\begin{equation}
\underline{\partial} \psi (x) = 0  \label{weyl1}
\end{equation}
shall serve now as a first Ansatz for a relativistically invariant wave equation.
In this context, relativistic invariance means that the wave equation (\ref{weyl1}) holds in all inertial systems,
a fact that is readily verified.
From $\underline{\partial} \psi(x) = 0$ trivially follows $A \underline{\partial} \psi(x) = 0$.
Here, $A$ is an $SL(2, \mathds{C})$-matrix associated with a Lorentz transformation $\Lambda$.
In wise foresight, one requests that $\psi(x)$ obeys the manifestly covariant transformation law in accordance with
eq. (\ref{lefttrafo})
\begin{equation}
\psi'(x') = \epsilon A^* \epsilon^{-1} \psi(\Lambda^{-1} x') \, , \quad \epsilon = i \sigma_2 =
\left(\begin{array}{cc}
0 & 1  \\
-1 & 0 \\
\end{array} \right) \, .
\end{equation}
The special property of the matrix $A \in SL(2, \mathds{C}) = Sp(2, \mathds{C})$
\begin{equation}
A^T \epsilon A = \epsilon
\end{equation}
implies
\begin{equation}
(A^T)^{-1} = \epsilon A \epsilon^{-1} \, , \quad (A^+)^{-1} = (A^{-1})^+ = \epsilon A^* \epsilon^{-1} \, ,
\end{equation}
hence
\begin{equation}
0 = A \underline{\partial} \psi(x) = A \underline{\partial} \underbrace{A^+ (A^+)^{-1}}_{\mathds{1}_2} \psi(x)
= \underbrace{A \underline{\partial} A^+}_{\underline{\partial}'} \underbrace{\epsilon A^* \epsilon^{-1} \psi(x)}_{\psi'(x')}
= \underline{\partial}' \psi' (x') \, . \label{left_weyl}
\end{equation}
Obviously, the 'natural law' $\underline{\partial} \psi=0$ is valid in a manifestly Lorentz-invariant form in every inertial system.
Manifest Lorentz invariance of a formalism provides great advantages from the calculational point of view,
but this certainly does not imply that more involved formalisms depending on a specific frame of reference
may play a r\^ole in theoretical physics.\\

Spinors transforming according to eq. (\ref{lefttrafo}) are called {\emph{left-chiral}} spinors.
One has to mention that this {\emph{chirality}} (from the greek $\chi \varepsilon \iota \rho$, 'hand')
should not be confused with the {\emph{helicity}} (from the greek $\varepsilon \lambda \iota \xi$, the 'twisted')
of the particles described by the wave functions discussed here.
Helicity is defined via the direction of the momentum and the angular momentum of a particle,
and it is not a Lorentz-invariant property for massive particles. In the massless case however,
helicity can be linked directly to chirality.\\

In the non-interacting case, a spinor $\psi_L(x)$ as in eq. (\ref{left_weyl}) obeys the so-called
{\emph{left-chiral Weyl equation}}
\begin{equation}
\underline{\partial} \psi_L (x) = (\sigma_0 \partial_0 - \vec{\sigma} \cdot \vec{\nabla}) \psi_L (x) = 0 \, ,
\end{equation}
which has been put in a form explicitly containing the nabla operator 
\begin{equation}
\vec{\nabla} =
\left(\begin{array}{c}
\partial/\partial x^1 \\
\partial/\partial x^2 \\
\partial/\partial x^3 \\
\end{array} \right)
=
- \left(\begin{array}{c}
\partial^1 \\
\partial^2\\
\partial^3 \\
\end{array} \right) \, .
\end{equation}
Of course, the {right-chiral} case is missing so far in the present discussion.
Therefore, one considers the operator
\begin{equation}
\overline{\partial} = \overline{\sigma}_\mu \partial^\mu = \epsilon \underline{\partial}^T \epsilon^{-1}
= \sigma_0 \partial_0 + \vec{\sigma} \cdot \vec{\nabla} = \overline{\sigma}_\mu^+ \partial^\mu=
\left(\begin{array}{cc}
\partial^0 - \partial^3 & -\partial^1 + i \partial^2  \\
-\partial^1 - i \partial^2 & \partial^0 + \partial^3   \\
\end{array} \right) \, ,
\end{equation}
where the easily verifiable identites $\epsilon \vec{\sigma} \epsilon^{-1} = - \vec{\sigma^{_T}}=-\vec{\sigma}^*$
have been used.
Then, the {\emph{right-chiral Weyl equation} for a right-chiral spinor $\psi_R$ reads
\begin{equation}
\overline{\partial} \psi_R (x) =  (\sigma_0 \partial_0 + \vec{\sigma} \cdot \vec{\nabla}) \psi_R (x) = 0 \, ,
\end{equation}
and again one can check for the manifest Lorentz invariance of this equation.
The transformation law (\ref{traflaw1}) implies for the operator $\overline{\partial}$ the transformation law
\begin{equation}
\overline{\partial}'= \epsilon [A \underline{\partial} A^+ ]^T \epsilon^{-1} =
\underbrace{\epsilon A^* \epsilon^{-1}}_{(A^+)^{-1}} \epsilon \underline{\partial}^T \epsilon^{-1} 
\underbrace{\epsilon A^T \epsilon^{-1}}_{A^{-1}} =(A^+)^{-1} \overline{\partial} A^{-1} \, ,
\end{equation}
and postulating the simple transformation law correponding to the right-chiral $(0, \frac{1}{2})$-representation
for the right-chiral field $\psi_R(x) $
\begin{equation}
\psi'(x') = A \psi(x) 
\end{equation}
directly leads to the desired result
\begin{equation}
\overline{\partial}' \psi' (x') = (A^+)^{-1} \overline{\partial} A^{-1} A \psi_R (x) = (A^+)^{-1} \overline{\partial}
\psi_R (x) = 0 \, .
\end{equation}\\
Together, the operators $\underline{\partial}$ and $\overline{\partial}$ possess the interesting property
\begin{displaymath}
\underline{\partial} \overline{\partial} =  \overline{\partial} \underline{\partial}  =
\left(\begin{array}{cc}
\partial^0 - \partial^3 & -\partial^1 + i \partial^2  \\
-\partial^1 - i \partial^2 & \partial^0 + \partial^3   \\
\end{array} \right) 
\left(\begin{array}{cc}
\partial^0 + \partial^3 & \partial^1 - i \partial^2  \\
\partial^1 + i \partial^2 & \partial^0 - \partial^3   \\
\end{array} \right) 
\end{displaymath}
\begin{equation}
=
\left(\begin{array}{cc}
\partial_0^2 -\partial_1^2 -\partial_2^2-\partial_3^2  & 0  \\
0  & \partial_0^2 -\partial_1^2 -\partial_2^2-\partial_3^2   \\
\end{array} \right) 
=
\left(\begin{array}{cc}
\Box  & 0  \\
0  & \Box \\
\end{array} \right) \, , \label{del_del}
\end{equation}
which can be expressed in a more elegant manner by exploiting the analytic symmetry
$\partial^\mu \partial^\nu = \partial^\nu \partial^\mu$ for functions of class $C^2$
\begin{equation}
\underline{\partial} \overline{\partial} = \sigma_\mu \partial^\mu \overline{\sigma}_\nu \partial^\nu
=\frac{1}{2} (\sigma_\mu \overline{\sigma}_\nu + \sigma_\nu \overline{\sigma}_\mu) \partial^\mu \partial^\nu=
g_{\mu \nu} \partial^\mu \partial^\nu \sigma_0 = \Box \mathds{1}_2 \, . \label{gordon_components}
\end{equation}
Eq. (\ref{gordon_components}) implies that both field components of a left- or right-chiral Weyl field
fulfill the Klein-Gordon equation.
From a group theoretical perspective, the differential operators
$\underline{\partial}$ and $\overline{\partial}$ are of a more fundamental significance that the
wave operator $\Box$, since the two two-component spinor operators,
which are related to the $(1/2,0)$- and $(0,1/2)$-representations of the proper orthochronous Lorentz group,
allow for the construction of wave equation for higher-spin fields with more involved transformation
properties. The Klein-Gordon wave operator, which is linked to the trivial representation of the Lorentz group,
does not contain this group theoretical information.\\

The Weyl equations do not describe a parity invariant world.
Introducing a passive parity transformation
\begin{equation}
\Lambda_P: \, (x^0, \vec{x}) \rightarrow (x'^0, \vec{x}') = (x^0, -\vec{x}) \, , \quad
SO^+(1,3) \not \ni \Lambda_P \in O(1,3) 
\end{equation}
and considering an observer describing the dynamics of a Weyl field by $\psi(x)$ and a point reflected observer
describing the same Weyl field in 'his own words' by $\psi'(x')$, one must have a linear transformation law
connecting the mathematical entities used by the two observers
\begin{equation}
\psi'(x') = A_P \psi(x) = A_P \psi(x'^0, -\vec{x}')  \label{parity_weyl}
\end{equation}
with an appropriate $2 \times 2$-matrix $A_P$ 
which makes it possible to translate theoretical or experimental aspects related to the Weyl field
from one observer to the other.
It is a simple exercise to show that no matrix $A_P$ exists such that
both $\psi(x)$ and $\psi'(x')$ fulfill the left-chiral (or right-chiral) Weyl equation at the same time.
In fact, a parity transformation transforms a left-chiral field into a right-chiral field and vice versa.
Of course one may wonder how it is possible to mirror an observer.
Anyway, it is much easier to boost or to rotate a person or a measuring device.\\

\subsection{Two-component Majorana equations}
The Weyl equations suffer from the disadvantage that they do not describe massive particles.
Modifying, e.g., the left-chiral Weyl equation by a naive mass term according to
\begin{equation}
\underline{\partial} \psi (x) + \hat{m} \psi (x) = 0 \, , \quad \hat{m} \in Mat(2, \mathds{C}) \backslash \mathds{O}_2
\label{weyl_mass}
\end{equation}
with an arbitrary but non-vanishing $2 \times 2$ mass matrix $\hat{m}$,
the wave equation turns out to be non-Lorentz invariant.
A solution $\psi'(x')$ of the left-chiral Weyl equation in an inertial system IS'
does not fulfill the Weyl equation in a different inertial system IS since
\begin{equation}
\underline{\partial}' \psi'(x') + \hat{m} \psi'(x') = A \underline{\partial} \psi(x) + \hat{m}
\epsilon A^* \epsilon^{-1} \psi(x) \neq 0 \, .
\end{equation}
In 1937, Ettore Majorana found an unconventional
way out of this disturbing situation by coupling a field with its complex conjugate field \cite{Majorana}.
The {\emph{left-chiral Majorana equation}}
\begin{equation}
\underline{\partial} \psi(x) + \eta m \epsilon^{-1} \psi (x)^* = 0 \, , \quad \eta \in U(1) \, , \quad m \in \mathds{R} \, ,
\end{equation}
obeys the desired transformation law,
\begin{equation}
\underline{\partial}' \psi'(x') + \eta m \epsilon^{-1} \psi'(x')^* = A \underline{\partial} \psi(x) + 
\eta m \epsilon^{-1} \underbrace{\epsilon A \epsilon^{-1} \psi(x)^*}_{\psi' (x')^*} = A ( \underline{\partial} \psi (x) +
\eta m \epsilon^{-1} \psi (x)^* ) = 0 \, . \label{ML}
\end{equation}
The mass term $m$ must be a scalar in order to commute with every possible spinor Lorentz transformation matrix
$A$ in eq. (\ref{ML}).\\

In complete analogy to the considerations above, one may write down the
{\emph{right-chiral Majorana equation}}.
Since the mass term can be equipped with a so-called Majorana phase in both the left- and the right-chiral case,
it is common usage in the literature to formulate the field equations and the corresponding transformation laws with
\begin{equation}
\epsilon^{-1} = - i \sigma_2 = \left(\begin{array}{cc}
0  & -1  \\
1  & 0 \\
\end{array} \right) 
\end{equation}
in the following manner ($m_{L,R} \in \mathds{R}$, $|\eta | =1$):
\begin{equation}
i \sigma_\mu \partial^\mu \psi_L(x) - \eta_L m_L (i \sigma_2) \psi_L (x)^* = 0 \, , \quad
\psi_L'(x') = \epsilon A^* \epsilon^{-1} \psi_L (\Lambda^{-1} x') \, ,
\end{equation}
\begin{equation}
i \overline{\sigma}_\mu \partial^\mu \psi_R(x) + \eta_R m_R (i \sigma_2) \psi_R (x)^* = 0 \, , \quad
\psi_R'(x') = A \psi_R (\Lambda^{-1} x') \, .
\end{equation}\\
Majorana fields play an important r\^ole as fundamental theoretical building blocks in supersymmetric quantum field
theories.\\

In the {\emph{non-interacting}} case the phases $\eta_{L,R}$ have no physical significance and can be removed by
a redefinition of the fields by the help of a global gauge transformation.
With $\psi_L(x) \rightarrow \psi'_L(x)=\psi_L(x) e^{-i \delta_L/2}$ and $\eta_L = e^{i \delta_L}$ one has, e.g., in the
left-chiral case
\begin{displaymath}
i (\sigma_0 \partial_0-\vec{\sigma} \cdot \vec{\nabla}) \psi_L(x)
-\eta_L m_L \epsilon \psi_L(x)^*=
i (\sigma_0 \partial_0-\vec{\sigma} \cdot \vec{\nabla}) e^{i \delta_L/2} \psi'_L(x)
-e^{i \delta_L} m_L \epsilon e^{-i \delta_L/2} \psi'_L (x)^*
\end{displaymath}
\begin{equation}
=e^{i \delta_L/2} [i (\sigma_0 \partial_0-\vec{\sigma} \cdot \vec{\nabla}) \psi'_L(x)
-m_L \epsilon \psi'_L(x)^*]=0 \, ,
\end{equation}
i.e. $\psi'_L (x)$ fulfills a phase-free Majorana equation.\\

For left-handed Majorana particles one obviously has due to $\vec{\nabla} \psi_L (x) =0$
\begin{equation}
i \dot{\psi}_{L,1} (x) =\eta_L m_L \psi_{L,2}^* (x) , \quad i \dot{\psi}_{L,2} (x) = -\eta_L m_L \psi_{L,1}^* (x) \, .
\end{equation}
Differentiating the left equation above with respect to time and using the complex conjugate equation at the right,
$\dot{\psi}_{L,2}^*  =-i \eta_L^* m_L \psi_{L,1}$, leads to
\begin{equation}
\ddot{\psi}_{L,1} (x) =-i \eta m_L \dot{\psi}_{L,2}^* (x) = -|\eta_L|^2 m_L^2 \psi_{L,1} (x)
=-m^2 \psi_{L,1} (x) \, .
\end{equation}
Therefore, $\psi_{L,1}$ is a linear combination of $e^{-imx^0}$- and
$e^{+imx^0}$-terms, and particles with their spin directed parallel or
anti-parallel to the $3$- or ($z$-)axis are described by the wave functions
($\psi_2 =-\frac{i}{m} \eta \dot{\psi}_1^*$)
\begin{equation}
\psi_{L,{+\frac{1}{2}}} (x) =\left(\begin{array}{c} 1 \\ 0  \end{array}\right)
e^{-imx^0}+
\eta \left(\begin{array}{c} 0 \\ 1  \end{array}\right) e^{+imx^0} \, , \label{majo_rest1}
\end{equation}
\begin{equation}
\psi_{L,{-\frac{1}{2}}} (x) =\left(\begin{array}{c} 0 \\ 1  \end{array}\right)
e^{-imx^0}+
\eta \left(\begin{array}{c} -1 \\ 0  \end{array}\right) e^{+imx^0} \, . \label{majo_rest2}
\end{equation}

The wave functions given by eqns. (\ref{majo_rest1}) and (\ref{majo_rest2})
can be sped up, e.g.,  by Wigner boosts.
Both the left- and the right-chiral Majorana fields $\psi_{L,R}$
describe one species of particles in the following sense: all {\emph{plane wave}} solutions
of the corresponding left- or right-chiral Majorana equations can be transformed
into each other by appropriate Poincar\'e transformations, i.e. by Lorentz transformations
and spacetime translations. Starting from the idealized, improper state of a particle at rest with a given
spin direction, all other states of the particle with sharp momentum can be generated by
boosts and rotations.

\section{Spin-$\frac{1}{2}$: Four-component complex and real spinor wave equations}
\subsection{Dirac equation}
A left-chiral two-component spin-$\frac{1}{2}$ field obeying the transformation law
$\psi'_L(x') = \epsilon A^* \epsilon^{-1} \psi_L(x)$
can be coupled to a right-chiral field with transformation law
$\psi'_R (x') = A \psi_R (x)$ - and vice versa -
with a coupling strength expressed by a mass term $m$ according to
\begin{equation}
(\sigma_0 \partial_0 - \vec{\sigma} \cdot \vec{\nabla}) \psi_L (x) + i m \psi_R (x) = 0 \, , \label{dirac_part_1}
\end{equation}
\begin{equation}
(\sigma_0 \partial_0 + \vec{\sigma} \cdot \vec{\nabla}) \psi_R (x) + i m \psi_L (x) = 0 \, . \label{dirac_part_2}
\end{equation}
The coupled system of equations (\ref{dirac_part_1}) and (\ref{dirac_part_2}) is manifestly Lorentz invariant, since
\begin{equation}
\underline{\partial}' \psi'_L (x') + im \psi'_R(x') = A \underline{\partial} \psi_L (x) + i m A \psi_R (x) =
A ( \underline{\partial} \psi_L (x) + i m  \psi_R (x) ) = 0 \, ,
\end{equation}
\begin{equation}
\overline{\partial}' \psi'_R (x') + im \psi'_L(x') = (A^+)^{-1} \overline{\partial} \psi_R (x) +
i m \epsilon A^* \epsilon^{-1} \psi_L (x) =
(A^+)^{-1} ( \overline{\partial} \psi_R (x) + i m  \psi_L (x) ) = 0 \, .
\end{equation}
Casting the eqns. (\ref{dirac_part_1}) and (\ref{dirac_part_2}) into the form
\begin{equation}
i \left(\begin{array}{cc}
0 & \sigma_0 \partial_0 - \vec{\sigma} \cdot \vec{\nabla} \\
\sigma_0 \partial_0 + \vec{\sigma} \cdot \vec{\nabla}  & 0 \\
\end{array}\right)
\left(\begin{array}{c}
\psi_R (x)\\
\psi_L (x)\\
\end{array}\right)
 -
\left(\begin{array}{cc}
m & 0 \\
0 & m \\
\end{array}\right)
\left(\begin{array}{c}
\psi_R (x)\\
\psi_L (x)\\
\end{array}\right) = 0
\end{equation}
and introducing the so-called {\emph{gamma matrices in chiral representation}}
\begin{equation}
\tilde{\gamma}^0 = \tilde{\gamma}_0 = \left(\begin{array}{cc}
0 & \mathds{1}_2 \\
\mathds{1}_2 & 0 \\
\end{array}\right) \, , \quad
\tilde{\gamma}^k = -\tilde{\gamma}_k = \left(\begin{array}{cc}
0 & -\sigma_k \\
\sigma_k & 0 \\
\end{array}\right) \, , \quad k= 1,2,3 \, ,
\end{equation}
one obtains the Dirac equation in its chiral representation 
\begin{equation}
i \tilde{\gamma}^\mu \partial_\mu \Psi (x) - m \Psi (x) = 0 \, ,
\end{equation}
with
\begin{equation}
\Psi(x) = \left(\begin{array}{c}
\psi_R (x)\\
\psi_L (x)\\
\end{array}\right) \, , \quad \tilde{\gamma}^\mu =
 \left(\begin{array}{cc}
0 & \sigma^\mu \\
\bar{\sigma}^\mu & 0 \\
\end{array}\right) \, .
\end{equation}
It is straightforward to check that the gamma matrices
fulfill the anti-commutation relations
\begin{equation}
\{ \tilde{\gamma}^\mu , \tilde{\gamma}^\nu \} = \tilde{\gamma}^\mu \tilde{\gamma}^\nu +
\tilde{\gamma}^\nu \tilde{\gamma}^\mu
= 2 g^{\mu \nu} \mathds{1}_4 \, ,
\label{dirac_anticomm}
\end{equation}
e.g., one has
\begin{equation}
(\tilde{\gamma}^1)^2 =
 \left(\begin{array}{cc}
0 & -\sigma_1 \\
\sigma_1 & 0 \\
\end{array}\right)
 \left(\begin{array}{cc}
0 & -\sigma_1 \\
\sigma_1 & 0 \\
\end{array}\right) =
 \left(\begin{array}{cc}
-\mathds{1}_2 & \mathds{O}_2 \\
\mathds{O}_2 & -\mathds{1}_2 \\
\end{array}\right) = -\mathds{1}_4 \, .
\end{equation}

Historically, the relations (\ref{dirac_anticomm}) were found in 1928 by Paul Adrien Maurice Dirac in his Ansatz \cite{Dirac}
to 'linearize' the Klein-Gordon equation according to
\begin{equation}
(\Box + m^2) \varphi(x) \, \rightarrow \, 
(-i \gamma^\nu \partial_\nu  \mp m ) ( i \gamma^\mu \partial_\mu  \mp m) \Psi (x) =
(\gamma^\nu \gamma^\mu \partial_\nu \partial_\mu + m^2) \Psi (x) = 0 \, ,
\end{equation}
which led him to conditions for the coefficients $\gamma^\mu= g^{\mu \nu} \gamma_\nu$
\begin{equation}
\displaystyle \Box = \gamma^\nu \gamma^\mu \partial_\nu \partial_\mu = \frac{1}{2}(\gamma^\mu \gamma^\nu +
\gamma^\nu \gamma^\mu) \partial_\mu \partial_\nu =\frac{1}{2}\{ \gamma^\mu, \gamma^\nu \} \partial_\mu \partial_\nu=
g^{\mu \nu} \partial_\mu \partial_\nu
\end{equation}
enforcing the introduction of a {\emph{Clifford algebra}} of gamma matrices $\gamma^0$, $\gamma^1$, $\gamma^2$
and $\gamma^3$, which can be represented in the lowest-dimensional case by $4 \times 4$-matrices.\\

It turns out that the gamma matrices can be represented in different ways.
The anti-commutation relations (\ref{dirac_anticomm}) are invariant with respect to a similarity transformation
with a non-singular matrix in $GL(4,\mathds{C})$, and apart from the chiral representation the literature
tends to use a standard representation with matrices $\gamma^\mu$ called the Dirac representation which is
linked to the chiral representation by
\begin{equation}
\tilde{\gamma}^\mu=U \gamma^\mu_{Dirac} U^{-1} \; , \quad
U=\frac{1}{\sqrt{2}} \left(\begin{array}{rr} \mathds{1}_2 & \mathds{1}_2 \\
\mathds{1}_2 & -\mathds{1}_2 \end{array}\right) \;\;\;, \quad U^{-1}=U^+ \, .
\label{dirac_basis_change}
\end{equation}
In the sequel, chiral gamma matrices shall be denoted by $\tilde{\gamma}^\mu$ and
(standard) Dirac matrices by $\gamma^\mu$.
The standard Dirac matrices are explicitly given by 
\begin{equation}
\gamma^0=\left(\begin{array}{rr} \mathds{1}_2 & 0 \\
0 & -\mathds{1}_2 \end{array}\right)
=\left(\begin{array}{rr} \sigma_0 & 0 \\
0 & -\sigma_0 \end{array}\right)
\, , \quad
\gamma^k=\left(\begin{array}{cc} 0 & \sigma_k \\
-\sigma_k & 0 \end{array}\right) \, ,
\end{equation}
and for many purposes, it is convenient to define a matrix $\gamma^5$ in a representation-independent
manner 
\begin{equation}
\gamma_5=\gamma^5=i \gamma^0 \gamma^1 \gamma^2 \gamma^3 \, .
\end{equation}
It is well-known that the solutions of the Dirac equation describe spin-$\frac{1}{2}$ particles together with their antiparticles
with the same mass. The Dirac matrices $\gamma^\mu$ are especially well-suited for investigations of the
low-energy limit of the Dirac equation.\\

The matrix $U$ in the eqns. (\ref{dirac_basis_change}) is unitary; as a matter of fact, all representations of the gamma matrices
which are unitarily equivalent to the chiral or Dirac representation exhibit the following
(anti-)\-Hermiticity relations
\begin{equation}
\gamma_0 \gamma_\mu \gamma_0 = \gamma_\mu^+ \, ,
\end{equation}
respectively
\begin{equation}
\gamma_0^+ = \gamma_0 \, , \quad \gamma_k^+ = - \gamma_k \, ,
\end{equation}
which provide some advantages for the discussion of energy and momentum observables.\\

An important result of the theory of Clifford algebras states that each set of four
$4 \times 4$-matrices fulfilling the anticommutation relations (\ref{dirac_anticomm})
can be brought into the chiral or Dirac form by a similarity transformation of the kind
(\ref{dirac_basis_change}), where $U$ is invertible but not necessarily unitary.
This nice feature enables theoretical physicists working
in different solar systems to compare their calculations by some simple conversions.
In this sense, the Dirac equation is universal.\\

Applying a similarity transformation to the Dirac matrices according to
\begin{equation}
\hat{\gamma}_\mu = B \gamma_\mu B^{-1} \, \, \rightarrow \, \,
\{ \hat{\gamma}_\mu , \hat{\gamma}_\nu \} = B \gamma_\mu B^{-1} B \gamma_\nu B^{-1}
+ B \gamma_\nu B^{-1} B \gamma_\mu B^{-1} = B \{ \gamma_\mu , \gamma_\nu \} B^{-1}
= 2 g_{\mu \nu} \mathds{1}_4 \label{aequiv_dirac}
\end{equation}
with an invertible matrix $B$, then the transformed Dirac spinor $\hat{\Psi} = B \Psi$ fulfills the
Dirac equation with the new gamma matrices $\hat{\gamma}^\mu$ again, since
from $(i \gamma_\mu \partial^\mu - m ) \Psi (x) = 0$ follows
\begin{equation}
(i \hat{\gamma}_\mu \partial^\mu -m) \hat{\Psi} (x) =
(i B \gamma_\mu B^{-1} \partial^\mu - m ) B \Psi (x) =
B ( i \gamma_\mu \partial^\mu - m ) \Psi(x) = 0 \, .
\end{equation}

From now on, spacetime arguments will be omitted for the sake of notational brevity.
In the eqns. (\ref{dirac_part_1}) and (\ref{dirac_part_2}), solely one single {\emph{real}} mass term
coupling a left- and a right-chiral two-component field shows up.
Indeed, a more general Ansatz
\begin{equation}
i \underline{\partial} \psi_L  - \tilde{m}_{D,R} \psi_R = 0 \, , \label{dirac_part_1_gen}
\end{equation}
\begin{equation}
i \overline{\partial} \psi_R  - \tilde{m}_{D,L} \psi_L = 0  \label{dirac_part_2_gen}
\end{equation}
with complex chiral mass terms $\tilde{m}_{D,R}$ and $\tilde{m}_{D,L}$ is conceivable.
Acting with the operator $-i \overline{\partial}$ on eq. (\ref{dirac_part_1_gen})
and using eq. (\ref{dirac_part_2_gen}) yields
\begin{equation}
\overline{\partial} \underline{\partial} \psi_L + i \tilde{m}_{D,R} \overline{\partial} \psi_R =
\Box \psi_L + \tilde{m}_{D,R} \tilde{m}_{D,L} \psi_L = 0 \, .
\end{equation}
Hence, the left-chiral part $\psi_L$ respects a Klein-Gordon-type equation, and the same
follows in complete analogy for the right-chiral part
$\Box \psi_R + \tilde{m}_{D,R} \tilde{m}_{D,L} \psi_R = 0$.
However, for the correct energy-momentum relation to hold true,
one must require
\begin{equation}
\tilde{m}_{D,R} \tilde{m}_{D,L} = m^2 \ge 0 \, .
\end{equation}
The degenerate case $m^2=0$ is not particularly interesting.
E.g., for
\begin{equation}
i \underline{\partial} \psi_L = 0 \, , \quad i \overline{\partial} \psi_R - \tilde{m} \psi_L = 0 
\end{equation}
with $\tilde{m} \neq 0$ follows $\Box \psi_R = \Box \psi_L =0$ and $\psi_L = i \tilde{m}^{-1}
\overline{\partial} \psi_R$, therefore $\psi_L$ is determined by the massless field $\psi_R$.
Writing the mass terms for $m^2 > 0$ in polar form
\begin{equation}
\tilde{m}_{D,R} = m_{D,R} e^{+i \varphi_{D,R}} \, , \quad
\tilde{m}_{D,L} = m_{D,L} e^{+i \varphi_{D,L}} = m_{D,L} e^{-i \varphi_{D,R}}= m_{D,L} e^{-i \varphi_D}
\end{equation}
with $m_{D,R}$, $m_{D,L} > 0$ and $\varphi_{D,R}= -\varphi_{D,L} = \varphi_D \in (\pi, \pi]$,
the Ans\"atze (\ref{dirac_part_1_gen}) and (\ref{dirac_part_2_gen}) can be written as
\begin{equation}
i \underline{\partial} \psi_L - \sqrt{m_{D,R} m_{D,L}} e^{+i \varphi_{D}} \sqrt{\frac{m_{D,R}}{m_{D,L}}} \psi_R = 0 \, ,
\end{equation}
\begin{equation}
i \overline{\partial} \sqrt{\frac{m_{D,R}}{m_{D,L}}}  \psi_R -
\sqrt{m_{D,L} m_{D,L}} e^{-i \varphi_{D}} \sqrt{\frac{m_{D,R}}{m_{D,L}}} \psi_L = 0 \, .
\end{equation}
Rescaling the right-chiral field
\begin{equation}
\psi'_R =  \sqrt{\frac{m_{D,R}}{m_{D,L}}}  \psi_R \label{rescaling}
\end{equation}
and introducing a Dirac mass term
$m_D=\sqrt{m_{D,R} m_{D,L}}$
finally yields the Dirac equation involving a single Dirac phase $\varphi_D$
\begin{equation}
i \underline{\partial} \psi_L - m_D e^{+i \varphi_{D}} \psi'_R = 0 \, ,
\end{equation}
\begin{equation}
i \overline{\partial} \psi'_R - m_D e^{-i \varphi_{D}} \psi_L = 0 \, .
\end{equation}
Only one real mass term $m_D$ is relevant for the present theory from the physical point of view.
Of course, one may argue about the physical relevance of parameters in non-interacting
theories. Eventually, the phase factors $e^{\pm i \varphi_D}$
can be trivially eliminated by a chiral phase transformation
\begin{equation}
\Gamma_L^\alpha: \psi_L \rightarrow \tilde{\psi}_L =\psi_L  e^{-i \alpha} \, , \quad \psi'_R \rightarrow \tilde{\psi}_R = \psi'_R \, ,
\label{chiral_left_phase}
\end{equation}
with $\alpha = \varphi_D$ or 
\begin{equation}
\Gamma_R^\beta: \psi_L \rightarrow \tilde{\psi}_L = \psi_L \, , \quad \psi'_R \rightarrow \tilde{\psi}_R = \psi'_R e^{-i \beta}
\label{chiral_right_phase}
\end{equation}
with $\beta=-\varphi_D$,
so that the fields $\tilde{\psi}_{L,R}$ fulfill the phase-free Dirac equation
\begin{equation}
i \underline{\partial} \tilde{\psi}_L (x) - m_D \tilde{\psi}_R = 0 \, ,
\end{equation}
\begin{equation}
i \overline{\partial} \tilde{\psi}_R (x) - m_D \tilde{\psi}_L = 0 \, .
\end{equation}
These phase transformations do {\emph{not}} represent a gauge transformation of the four-component
Dirac spinor, merely one has to state that the same physical information is encoded in the fields
$\tilde{\psi}_{L,R}$ as in $\psi_L$ and $\psi'_R$. Thus, the transformation trick above
does not imply that phases in interacting theories are not related to measurable quantities. 
A \emph{gauge transformation}
\begin{equation}
\Gamma^\alpha = \Gamma_L^\alpha \Gamma_R^{\alpha} : \psi_{L,R} \rightarrow e^{-i \alpha} \psi_{L,R}
\end{equation}
would leave $e^{i \varphi_D}$ unchanged.\\

Actually, purely imaginary representations of the gamma matrices which are unitarily equivalent to the standard Dirac matrices
exist. Using such matrices in a so-called Majorana representation, the Dirac equation becomes a purely real differential equation.

\subsection{Real four-component Majorana equation}
Decomposing the two complex components of a left-chiral spinor according to
\begin{equation}
\psi_L =  \left(\begin{array}{c}
\psi_1 \\ \psi_2
\end{array}\right)
=
\left(\begin{array}{c}
\Psi_1 + i \Psi_2 \\ \Psi_3 + i \Psi_4
\end{array}\right) \, , \quad \Psi_{1,2,3,4} \in \mathds{R} \, , \label{decomp_majo}
\end{equation}
one obtains from the two-component Majorana equation
(for the sake of simplicity a trivial Majorana phase such that $\eta_L=1$ shall be used for
the forthcoming considerations)
\begin{equation}
i \sigma_\mu \partial^\mu \psi_L  - m(i \sigma_2) \psi_L^* = 
i \left(\begin{array}{cc}
\partial^0+ \partial^3 & \partial^1 - i \partial^2 \\
\partial^1 + i \partial^2 & \partial^0 - \partial^3
\end{array}\right)
\left(\begin{array}{c}
\Psi_1 + i \Psi_2 \\ \Psi_3 + i \Psi_4
\end{array}\right)
- i m
\left(\begin{array}{c}
-i \Psi_3 - \Psi_4 \\ \ +i \Psi_1 +  \Psi_2
\end{array}\right) = 0
\end{equation}
after a separation into real and imaginary parts, the real linear system of first order differential equations
\begin{eqnarray}
\partial^1 \Psi_2 + \partial^2 \Psi_1 + \partial^0 \Psi_4 - \partial^3 \Psi_4 & - & m \Psi_1 = 0 \, ,\\
\partial^1 \Psi_1 - \partial^2 \Psi_2 + \partial^0 \Psi_3  - \partial^3 \Psi_3 & - & m \Psi_2 = 0 \, ,\\
\partial^0 \Psi_2  + \partial^3 \Psi_2 - \partial^2 \Psi_3 + \partial^1 \Psi_4 & + & m \Psi_3 = 0 \, ,\\
\partial^0 \Psi_1 + \partial^3 \Psi_1 + \partial^1 \Psi_3 + \partial^2 \Psi_4  & + & m \Psi_4 = 0 \, ,
\end{eqnarray}
or
\begin{equation}
\left(\begin{array}{cccc}
\partial^2 & \partial^1 & 0 & \partial^0 - \partial^3 \\
\partial^1 & - \partial^2 & \partial^0 - \partial^3 & 0 \\
0 & -\partial^0 - \partial^3 & \partial^2 & - \partial^1 \\
- \partial^0 - \partial^3 & 0 & - \partial^1 & - \partial^2
\end{array}\right)
\left(\begin{array}{c}
\Psi_1 \\ \Psi_2 \\ \Psi_3 \\ \Psi_4
\end{array}\right) -
m 
\left(\begin{array}{c}
\Psi_1 \\ \Psi_2 \\ \Psi_3 \\ \Psi_4
\end{array}\right)
= 0 \, . \label{majorana_vierer_reell}
\end{equation}\\
By the help of the purely imaginary Majorana (gamma) matrices
($\gamma_\mu^M = g_{\mu \nu} \gamma_M^\nu$)
\begin{displaymath}
\gamma_0^M =
\left(\begin{array}{cccc}
0 & 0 & 0 & -i \\
0 & 0 & -i & 0 \\
0 & i & 0 & 0 \\
i & 0 & 0 & 0
\end{array}\right) \, , \quad
\gamma_1^M =
\left(\begin{array}{cccc}
0 & -i & 0 & 0 \\
-i & 0 & 0 & 0 \\
0 & 0 & 0 & i \\
0 & 0 & i & 0
\end{array}\right) \, ,
\end{displaymath}
\begin{equation}
\gamma_2^M =
\left(\begin{array}{cccc}
-i & 0 & 0 & 0 \\
0 & i & 0 & 0 \\
0 & 0 & -i & 0 \\
0 & 0 & 0 & i
\end{array}\right) \, , \quad
\gamma_3^M =
\left(\begin{array}{cccc}
0 & 0 & 0 & i \\
0 & 0 & i & 0 \\
0 & i & 0 & 0 \\
i & 0 & 0 & 0
\end{array}\right) \, ,  \label{Majo_Dirac_Darst}
\end{equation}
eq. (\ref{majorana_vierer_reell}) can be cast into the Dirac equation form
$(i \gamma^{M}_\mu \partial^\mu - m ) \Psi_{M} = 0$ with a Majorana spinor $\Psi_{M}=(\Psi_1, \Psi_2, \Psi_3, \Psi_4)^T$,
rendering the Dirac equation a real differential equation.
It is readily verified that the $\gamma^{M}$-Majorana matrices in the representation (\ref{Majo_Dirac_Darst}) above
obey the mandatory anticommutation relations
$\{ \gamma^{M}_\mu, \gamma^{M}_\nu \} = 2 g_{\mu \nu} \mathds{1}_4$.\\

Furthermore, restricting the spinor components of $\Psi_{M}$ according to the original construction premised by eq.
(\ref{decomp_majo}) to real values only, the four-component Majorana equation
 $(i \gamma^{M}_\mu \partial^\mu - m ) \Psi_{M} = 0$ is completely equivalent to the two-component Majorana equation, and again
it describes (after second quantization) the dynamics of neutral spin-$\frac{1}{2}$ particles. However, abandoning the
requirement $\Psi_{M}=\Psi_{M}^*$, the number of the degrees of freedom described by the four-spinor doubles and
one is lead back to the theory describing two spin-$\frac{1}{2}$-(anti)particles through the Dirac equation.\\

A further, purely imaginary representation of the Majorana matrices spread in the literature is given by
\begin{displaymath}
\tilde{\gamma}^0_{M} = \left(\begin{array}{cccc}
0 & 0 & 0 & -i \\
0 & 0 & i & 0 \\
0 & -i & 0 & 0 \\
i & 0 & 0 & 0
\end{array}\right) =
\left(\begin{array}{cccc}
0 & \sigma_2\\
\sigma_2 & 0 
\end{array}\right)  , \, \,
\tilde{\gamma}^1_{M} = \left(\begin{array}{cccc}
i & 0 & 0 & 0 \\
0 & -i & 0 & 0 \\
0 & 0 & i & 0 \\
0 & 0 & 0 & -i
\end{array}\right) = i
\left(\begin{array}{cccc}
\sigma_3 & 0\\
0 & \sigma_3 
\end{array}\right)  ,
\end{displaymath}
\begin{equation}
\tilde{\gamma}^2_{M} = \left(\begin{array}{cccc}
0 & 0 & 0 & i \\
0 & 0 & -i & 0 \\
0 & -i & 0 & 0 \\
i & 0 & 0 & 0
\end{array}\right) =
\left(\begin{array}{cccc}
0 & -\sigma_2\\
\sigma_2 & 0 
\end{array}\right)  , \, \,
\tilde{\gamma}^3_{M} = \left(\begin{array}{cccc}
0 & -i & 0 & 0 \\
-i & 0 & 0 & 0 \\
0 & 0 & 0 & -i \\
0 & 0 & -i & 0
\end{array}\right) = -i
\left(\begin{array}{cccc}
\sigma_1 & 0\\
0 & \sigma_1 
\end{array}\right)  .
\end{equation}
This representation can be obtained from the original representation
(\ref{Majo_Dirac_Darst}) by the unitary transformation
\begin{equation}
\tilde{\gamma}^\mu_{M}= \tilde{U} \gamma^\mu_M \tilde{U}^{-1} \, , \quad
\tilde{U} = \frac{1}{2} \left(\begin{array}{cccc}
+1 & +1 & +1 & -1 \\
-1 & +1 & +1 & +1 \\
+1 & +1 & -1 & +1 \\
+1 & -1 & +1 & +1
\end{array}\right) \in O(4) \subset U(4)
\end{equation}
where $\det (\tilde{U}) = -1$.\\

The four-component spinor appearing in eq. (\ref{majorana_vierer_reell}) is real by definition, a fact which can be expressed
by the condition $\Psi_M = \Psi_M^*$. Therefore, complex conjugation can be interpreted as a charge conjugation operator
and the condition $\Psi_M^* = \Psi_M$ simply expresses the fact that a neutral Majorana particle is invariant under charge conjugation.
Applying a unitary similarity transformation on the Majorana matrices and the Majorana spinor according to eq. (\ref{aequiv_dirac})
\begin{equation}
\hat{\gamma}_\mu^M = U \gamma_\mu^M U^{-1} \,, \quad
\hat{\Psi}_M = U \Psi_M \, , \quad U^{-1}=U^+=(U^T)^* \, ,
\end{equation}
the condition that the Majorana-Dirac equation should describe neutral particles becomes
\begin{equation}
\Psi_M^* = (U^{-1} \hat{\Psi}_M)^* = U^T \hat{\Psi}_M^* = U^{-1} \hat{\Psi}_M = \Psi_M \, ,
\end{equation}
therefore the neutrality condition for the transformed four-components Majorana spinor now reads
\begin{equation}
\hat{\Psi}_M = U U^T \hat{\Psi}_M^* \, .
\end{equation}
For real, i.e. orthogonal $U \in O(4) \subset U(4)$ one has $U U^T = \mathds{1}_4$, and so again
$\hat{\Psi}_M^* = \hat{\Psi}_M$.\\

The discussion above illustrates the complete equivalence of the four-com\-po\-nent and the two-com\-po\-nent
Majorana formalism in the literature. The four-component field is related to a irreducible four-dimensional real
spinor representation of the Lorentz group, whereas the two-component formalism is based on the two
fundamental complex spinor representations.

\section{Weyl-Majorana-Dirac formalism}
Considering now the most general free field case, the left- and right-chiral fields can be coupled
via linear and anti-linear terms according to the following "Weyl-Majorana-Dirac equation"
\begin{equation}
i \sigma_\mu \partial^\mu \psi_L  - \eta_{{D,R}} m_{{D,R}} \psi_R - \eta_{{L}} m_{{L}} (i \sigma_2) \psi_L^* = 0 \, ,
\label{L}
\end{equation}
\begin{equation}
i \bar{\sigma}_\mu \partial^\mu \psi_R  - \eta_{{D,L}} m_{{D,L}} \psi_L + \eta_{{R}} m_{{R}} (i \sigma_2) \psi_R^* = 0 \, ,
\label{R}
\end{equation}
with non-negative mass terms $m_L$, $m_R$, $m_{D,L}$, and $m_{D,R}$, and phase terms $\eta_L$, $\eta_R$, $\eta_{D,L}$, and
$\eta_{D,R}$ in the unitary group $U(1)$. A polar decomposition of the complex mass terms according to
\begin{displaymath}
\tilde{m}_{D,L}= \eta_{D,L} m_{D,L} = m_{D,L} e^{i \varphi_{D,L}} \, , \quad
\tilde{m}_{D,R}= \eta_{D,R} m_{D,R} = m_{D,R} e^{i \varphi_{D,R}} \, ,
\end{displaymath}
\begin{equation}
\quad \tilde{m}_L=\eta_L m_L = m_L e^{i \varphi_L} \, , \quad
\tilde{m}_R=\eta_R m_R = m_R e^{i \varphi_R} 
\end{equation}
with phases $\varphi_{D,L}$, $\varphi_{D,R}$, $\varphi_{L}$, $\varphi_R \in (-\pi, \pi]$ can also be used for notational convenience.
When all mass terms vanish, the eqns. (\ref{L}) and (\ref{R}) trivially describe a left- and a right-chiral field.
But in the following, we consider the non-trivial Majorana-Dirac case where none of the mass terms above vanishes.\\

One may note first that
using $(i \sigma_2) \vec{\sigma}^* = - \vec{\sigma} (i \sigma_2)$ leads to
$(i \sigma_2) K {\bar{\sigma}}_\mu \partial^\mu = \sigma_\mu \partial^\mu (i \sigma_2) K$ or
\begin{equation}
\epsilon K \underline{\partial} = \overline{\partial} \epsilon K \, , \quad 
\epsilon K \overline{\partial} = \underline{\partial} \epsilon K \, ,
\end{equation}
where the operator $K$ denotes complex conjugation.
Hence, eqns. (\ref{L}) and (\ref{R}) are fully equivalent to
(keeping in mind that $(i \sigma_2)^2= \epsilon^2= -\mathds{1}_2$, $K \eta \varphi =\eta^* K \varphi$, and $Ki=-iK$)
\begin{equation}
i \overline{\partial} (\epsilon \psi^*_L) + \tilde{m}^*_{D,R} (\epsilon \psi^*_R) - \tilde{m}^*_L \psi_L = 0 \, , 
\end{equation}
\begin{equation}
i \underline{\partial} ( \epsilon \psi^*_R) + \tilde{m}^*_{D,L} (\epsilon \psi^*_L) + \tilde{m}^*_R \psi_R = 0 \, ,
\end{equation}
i.e., the left-chiral field $\psi_L$ is physically equivalent to a right-chiral field $\epsilon \psi^*_L$, whereas
the
 right-chiral field $\psi_R$ is equivalent to the left-chiral field $\epsilon \psi^*_R$.\\

Working with the original eqns. (\ref{L}) and (\ref{R}), which can be cast into the form
\begin{equation}
i
\left(\begin{array}{cc}
0 & \underline{\partial}  \\
\overline{\partial} & 0   \\
\end{array} \right) 
\left(\begin{array}{c}
\psi_R   \\
\psi_L    \\
\end{array} \right)
-
\left(\begin{array}{cc}
\tilde{m}_{D,R} & \tilde{m}_L \epsilon K \\
-\tilde{m}_R \epsilon K & \tilde{m}_{D,L}  \\
\end{array} \right)
\left(\begin{array}{c}
\psi_R   \\
\psi_L    \\
\end{array} \right)
= 0
\label{Majorana_Dirac}
\end{equation}
yields the compact representation
\begin{equation}
i \gamma_\mu \partial^\mu \Psi - \hat{m} \Psi = 0 \, , \quad
\Psi =
\left(\begin{array}{c}
\psi_R   \\
\psi_L    \\
\end{array} \right) \, ,
\label{wmd}
\end{equation}
with chiral Dirac matrices fulfilling the usual anti-commutation relations
$\{ \gamma_\mu , \gamma_\nu \} = 2 g_{\mu \nu} \mathds{1}_4$
and a non-linear mass operator $\hat{m}$.\\

Defining a dual mass operator
\begin{equation}
\check{m}=
\left(\begin{array}{cc}
\tilde{m}_{D,R} & \tilde{m}_R \epsilon K \\
-\tilde{m}_L \epsilon K & \tilde{m}_{D,L}  \\
\end{array} \right)
\end{equation}
one obtains
\begin{equation}
\check{m} \hat{m} =
\left(\begin{array}{cc}
\tilde{m}_{D,R}^2 + m_R^2 & (\tilde{m}_{D,R} \tilde{m}_L  + \tilde{m}^*_{D,L} \tilde{m}_{R}) \epsilon K\\
- (\tilde{m}^*_{D,R} \tilde{m}_L  + \tilde{m}_{D,L} \tilde{m}_R) \epsilon K & \tilde{m}_{D,L}^2 + m_L^2\\
\end{array} \right) \, .
\end{equation}
Acting with $(-i \gamma_\nu \partial^\nu - \check{m})$ on eq. (\ref{wmd}) leads to
\begin{equation}
(-i \gamma_\nu \partial^\nu - \check{m})(i \gamma_\mu \partial^\mu - \hat{m}) \Psi = 
(\Box + \check{m} \hat{m} + (i \gamma_\nu \partial^\nu)  \hat{m} - \check{m} (i \gamma_\mu \partial^\mu)) \Psi =
0 \, , 
\end{equation}
where (beware that $Ki=-iK$)
\begin{displaymath}
 (i \gamma_\nu \partial^\nu)  \hat{m} - \check{m} (i \gamma_\mu \partial^\mu)
\end{displaymath}
\begin{equation}
= i \left(\begin{array}{cc}
- \underline{\partial} \tilde{m}_R \epsilon K + \tilde{m}_R \epsilon K \overline{\partial} & 
( \tilde{m}_{D,L}  - \tilde{m}_{D,R}) \underline{\partial} \\
( \tilde{m}_{D,R}  - \tilde{m}_{D,L}) \overline{\partial}  & 
\overline{\partial} \tilde{m}_L \epsilon K - \tilde{m}_L \epsilon K \underline{\partial} \\
\end{array} \right) \, .
\end{equation}
Using $\epsilon K \underline{\partial} = \overline{\partial} \epsilon K$ and 
$\epsilon K \overline{\partial} = \underline{\partial} \epsilon K$ again,
the expression above reduces to
\begin{equation}
 (i \gamma_\nu \partial^\nu)  \hat{m} - \check{m} (i \gamma_\mu \partial^\mu)
= i \left(\begin{array}{cc}
0 & ( \tilde{m}_{D,L}  - \tilde{m}_{D,R}) \underline{\partial} \\
( \tilde{m}_{D,R}  - \tilde{m}_{D,L}) \overline{\partial} & 0 \\
\end{array} \right) \, .
\label{first_order}
\end{equation}
We remember now that it is in fact possible to rescale, e.g., the right-chiral field $\psi_R$ according to
eq. (\ref{rescaling}) in order to obtain field equations where the modulus of the Dirac mass terms fulfills
$m_{D,R} = m_{D,L}$.
Additionally, introducing a phase-transformed left-chiral field $\psi'_L$ according to eq. (\ref{chiral_left_phase})
\begin{equation}
\psi'_L= \Gamma^{\alpha/2}_L \psi_L = \psi_L e^{-i \alpha/2} \, ,
\end{equation}
eqns. (\ref{L}) and (\ref{R}) can be written as
\begin{equation}
i \sigma_\mu \partial^\mu \psi'_L  - \tilde{m}'_{{D,R}} \psi_R - \tilde{m}'_L (i \sigma_2) \psi'^*_L = 0 \, ,
\label{L_mod}
\end{equation}
\begin{equation}
i \bar{\sigma}_\mu \partial^\mu \psi_R   - \tilde{m}'_{D,L} \psi'_L  + \tilde{m}'_R (i \sigma_2) \psi_R^*  = 0 \, ,
\label{R_mod}
\end{equation}
with
\begin{equation}
\tilde{m}'_{D,L} = \tilde{m}_{D,L} e^{+i \alpha/2} \, , \quad \tilde{m}'_{D,R} = \tilde{m}_{D,R} e^{-i \alpha/2} \, , \quad
\tilde{m}'_L = \tilde{m}_L e^{-i \alpha} \, , \quad \tilde{m}'_R = \tilde{m}_R \, .
\end{equation}
A phase transformation of the right-chiral field $\psi_R$ only
\begin{equation}
\psi'_R= \Gamma^{\beta/2}_R \psi_R = \psi_R e^{-i \beta/2}
\end{equation}
modifies the mass parameters in eqns. (\ref{L}) and (\ref{R}) according to
\begin{equation}
\tilde{m}'_{D,L} = \tilde{m}_{D,L} e^{-i \beta/2} \, , \quad \tilde{m}'_{D,R} = \tilde{m}_{D,R} e^{+i \beta/2} \, , \quad
\tilde{m}'_R = \tilde{m}_R e^{-i \beta} \, , \quad \tilde{m}'_L = \tilde{m}_L \, .
\end{equation}
Observing that the left-chiral Dirac mass term $\tilde{m}_{D,L}$ picks up the opposite phase compared to the
right-chiral mass term $\tilde{m}_{D,R}$ under a phase transformation
and considering the effect of rescaling one of the fields $\psi_{L,R}$
shows that one could also start with an equivalent field theory where $\tilde{m}_{D,R} = \tilde{m}_{D,L} = \tilde{m}_D \neq 0$.
Additionally, the phase of $\tilde{m}_D$ can be chosen to fulfill
\begin{equation}
Re(\tilde{m}_D) \ge 0 \, . \label{phaseconv_dirac}
\end{equation}
For $\tilde{m}_D=\tilde{m}_{D,R} = \tilde{m}_{D,L}$, the operator $(i \gamma_\nu \partial^\nu)  \hat{m} - \check{m}
(i \gamma_\mu \partial^\mu)$ in eq. (\ref{first_order})
vanishes and the Majorana-Dirac equation (\ref{Majorana_Dirac}) is equivalent, after appropriate rescaling and phase transformation
of the corresponding fields, to the field equation
\begin{equation}
i
\left(\begin{array}{cc}
0 & \underline{\partial}  \\
\overline{\partial} & 0   \\
\end{array} \right) 
\left(\begin{array}{c}
\psi_R   \\
\psi_L    \\
\end{array} \right)
-
\left(\begin{array}{cc}
\tilde{m}_{D} & \tilde{m}_L \epsilon K \\
-\tilde{m}_R \epsilon K & \tilde{m}_{D}  \\
\end{array} \right)
\left(\begin{array}{c}
\psi_R   \\
\psi_L    \\
\end{array} \right)
= 0 \, ,
\end{equation}
where any superscripts due to aforegoing scaling and phase transformations have been omitted,
and the fields obey the Klein-Gordon equation with generalized mass terms
\begin{equation}
\Box
\left(\begin{array}{c}
\psi_R   \\
\psi_L    \\
\end{array} \right)
+
\left(\begin{array}{cc}
\tilde{m}_{D}^2 + m_R^2 & (\tilde{m}_{D} \tilde{m}_L  + \tilde{m}^*_{D} \tilde{m}_{R}) \epsilon K\\
- (\tilde{m}^*_{D} \tilde{m}_L  + \tilde{m}_{D} \tilde{m}_R) \epsilon K & \tilde{m}_{D}^2 + m_L^2\\
\end{array} \right)
\left(\begin{array}{c}
\psi_R   \\
\psi_L    \\
\end{array} \right) = 0
\, .
\end{equation}
However, there is still the freedom to perform a gauge transformation leaving the Dirac mass $\tilde{m}_D$
invariant but changing $\tilde{m}_K$ and $\tilde{m}_R$ by a common phase. This freedom can be used to
redefine the fields and correspondingly 
rotate the phases of $\tilde{m}_L$ and $\tilde{m}_R$ in order to obtain
\begin{equation}
\kappa := \tilde{m}_{D} \tilde{m}_L  + \tilde{m}^*_{D} \tilde{m}_{R}  \ge 0 \, ,
\end{equation}
and every Majorana-Dirac equation with suitably redefined fields leads to the Klein-Gordon equation
\begin{equation}
\Box
\left(\begin{array}{c}
\psi_R   \\
\psi_L    \\
\end{array} \right)
+
\left(\begin{array}{cc}
\tilde{m}_{D}^2 + m_R^2 &\kappa \epsilon K\\
- \tilde{\mu} \epsilon K & \tilde{m}_{D}^2 + m_L^2\\
\end{array} \right)
\left(\begin{array}{c}
\psi_R   \\
\psi_L    \\
\end{array} \right) = 0
\end{equation}
with $\tilde{\mu}=\tilde{m}^*_{D} \tilde{m}_L  + \tilde{m}_{D} \tilde{m}_R$.
This Klein-Gordon equation can be written in a manifestly {\emph{real}} form by introducing the
real spinor
\begin{equation}
\Phi = (\psi_1, \psi_2, \ldots \psi_8)^T \, ,
\end{equation}
with eight real components given by
\begin{equation}
\psi_R=
\left(\begin{array}{c}
\psi_1+ i \psi_2   \\
\psi_3 + i \psi_4    \\
\end{array} \right) \, , \quad 
\psi_L=
\left(\begin{array}{c}
\psi_5+ i \psi_6   \\
\psi_7 + i \psi_8    \\
\end{array} \right) 
\end{equation}
according to
\begin{equation}
\Box \Phi + \hat{M}^2 \Phi = 0 \, .
\label{real_kg}
\end{equation}
Obviously, the mass operator $\hat{M}^2$ must be {\emph{positive semi-definite}} in order to exclude time-asymmetric complex
mass solutions or even tachyonic solutions of the Majorana-Dirac equation, and it must be Hermitian in order to generate a unitary
dynamics of the single particle states described by the wave function $\Phi$.
Only then the solutions of eq. (\ref{real_kg}) describe well-behaved normalizable single particle states as part of a stable
theory which are eigenstates of the
energy-momentum squared Casimir operator of the double covering group of the inhomogeneous
Poincar\'e group $\bar{\mathcal{P}}_+^\uparrow
\simeq T_{1,3} \rtimes SL(2, \mathds{C})$, which is the semi-direct product of the time-space translation
group $T_{1,3}$ and the universal cover $SL(2, \mathds{C})$ of the proper orthochronous
Lorentz group $SO^+ (1,3)$. This condition restricts the admissible mass terms, as discussed in the following.\\

Using the abbreviations $\tilde{\mu}=\mu_1 + i \mu_2$ and
$\tilde{m}_{D}^2=\nu_1 + i \nu_2$, the
mass operator $\hat{M}^2$ reads
\begin{equation}
\hat{M}^2=
\left(\begin{array}{cccccccc}
\nu_1+m_R^2 & -\nu_2 & 0 & 0 & 0 & 0 & \kappa & 0 \\
\nu_2 & \nu_1+m_{R}^2 & 0 & 0 & 0 & 0 & 0 & -\kappa \\
0 & 0 & \nu_1+ m_R^2  & -\nu_2 & -\kappa & 0 & 0 & 0 \\
0 & 0 & \nu_2 & \nu_1+m_{R}^2 & 0 & \kappa & 0 & 0 \\
0 & 0 & -\mu_1 & -\mu_2 & \nu_1+m_{L}^2 & -\nu_2 & 0 & 0 \\
0 & 0 & -\mu_2 & \mu_1 & \nu_2 &\nu_1+m_{L}^2 &  0 & 0 \\
\mu_1 & \mu_2 & 0 & 0 & 0 & 0 & \nu_1+m_{L}^2 & -\nu_2 \\
\mu_2 & -\mu_1 & 0 & 0 & 0 & 0 & \nu_2 & \nu_1+m_{L}^2 \\
\end{array} \right) \, .
\end{equation}
Some straightforward algebra results in the following four doubly degenerate eigenvalues of $\hat{M}^2$
\begin{equation}
\lambda_{1,2}=\nu_1 + \frac{m_L^2+m_R^2}{2} \pm \sqrt{\biggl( \frac{m_L^2-m_R^2}{2} \biggr)^2 +
\kappa \tilde{\mu}-\nu_2^2+ i \nu_2 (m_L^2-m_R^2) } \, ,
\label{mass_complex_1}
\end{equation}
 \begin{equation}
\lambda_{3,4}=\nu_1 + \frac{m_L^2+m_R^2}{2} \pm \sqrt{\biggl( \frac{m_L^2-m_R^2}{2} \biggr)^2 +
\kappa \tilde{\mu}^*-\nu_2^2- i \nu_2 (m_L^2-m_R^2) } \, .
\label{mass_complex_2}
\end{equation}
The Hermiticity of $\hat{M}^2$ implies the conditions $\nu_2 = 0$ and $\mu_2=0$.
From $\nu_2$ follows that $\tilde{m}_D^2$ is real; considering additionally that $\kappa$ is real
and also $\tilde{\mu}$ must be real, $\tilde{m}_D$ becomes a real and positive parameter with
relation (\ref{phaseconv_dirac}).
Then, the eigenvalues of $\hat{M}^2$ become
\begin{equation}
\lambda_{\pm}=m_D^2 + \frac{m_L^2+m_R^2}{2} \pm \sqrt{\biggl( \frac{m_L^2-m_R^2}{2} \biggr)^2 +
m_D^2 (\tilde{m}_L + \tilde{m}_R)^2 } \, ,
\label{mass_complex_mod}
\end{equation}
with $m= \tilde{m}_L + \tilde{m}_R$ real. This is the origin of the Majorana phase $\varphi_M$ graphically
depicted in Fig. (\ref{fig_majorana_phase}).
Writing
\begin{equation}
\tilde{m}_R = m_R e^{i \varphi_R} \, , \quad \tilde{m}_L = m_L e^{i \varphi_L} \, , \quad 
\varphi_L = \varphi_R + \varphi_M \, ,
\end{equation}
the sine and the cosine law imply
\begin{equation}
m^2 = (\tilde{m}_L + \tilde{m}_R)^2 = m_R^2 + m_L^2 + 2 m_R m_L \cos \varphi_M
\end{equation}
together with
\begin{equation}
\sin \varphi_R = \frac{m_L}{m} \sin \varphi_M \, , \quad
\sin \varphi_L = - \frac{m_R}{m} \sin \varphi_M \, , \quad
\frac{\sin \varphi_L}{\sin \varphi_R} = - \frac{m_R}{m_L} \, .
\end{equation}
\begin{figure}[htb]
\begin{center}
\includegraphics[width=7.5cm, angle=270]{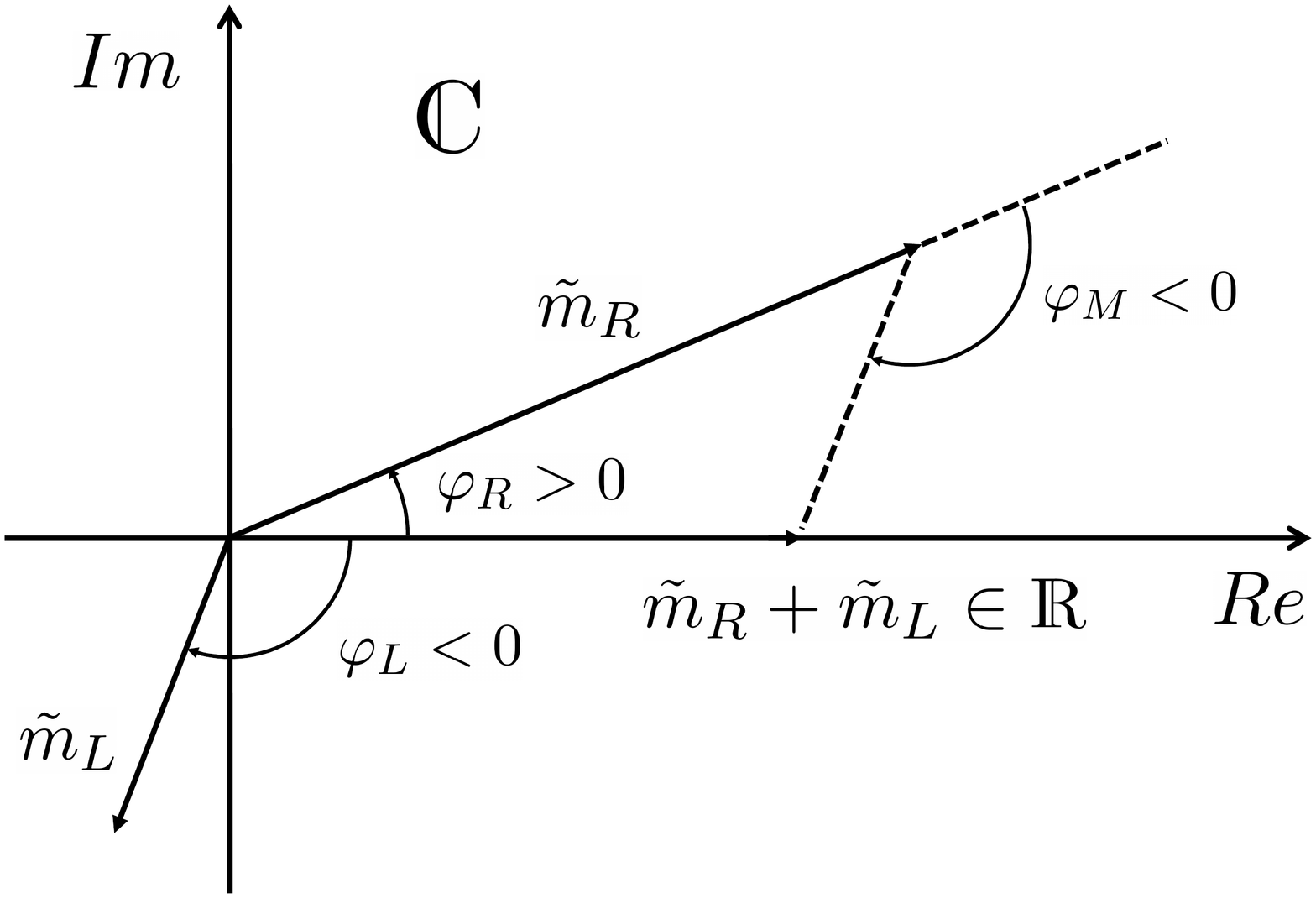}
\caption{Left- and right-chiral mass terms $\tilde{m}_{L,R}$ with a relative Majorana phase $\varphi_M$ adding up to
a real mass term $m=\tilde{m}_R+\tilde{m}_L$.}
\label{fig_majorana_phase}
\end{center}
\end{figure}
For the special case $\varphi_M = 0$, the two Majorana masses become
\begin{equation}
m_{\pm}^2=\Biggl[ \frac{m_L+m_R}{2} \pm \sqrt{\biggl( \frac{m_L-m_R}{2} \biggr)^2 + m_D^2} \,  \Biggr]^2 \, .
\end{equation} 
The explicit expressions for the mass eigenvalues illustrate how the presence of Majorana mass term split a
Dirac field into a couple of two Majorana fields.\\

The discussion of the rather trivial cases where one or several of the Majorana or Dirac mass terms are absent
or degenerate cases where some masses have the same modulus and special phase relations
is left to the reader as an interesting exercise.

\section{Finite-dimensional irreducible ray representations of the proper ortho\-chro\-nous Lorentz group}
\subsection{Complex representation theory of the $SL(2, \mathds{C})$}
In order to understand the findings of the last section from a more general perspective,
we finally leave the restricted framework of the fundamental representations of the proper orthochronous
Lorentz group and  shortly revisit the most important results from the theory of its real and
complex finite-dimensional representations \cite{Sexl}.
Such a discussion fits nicely into the considerations exposed so far for spin-$\frac{1}{2}$ fields only and will clarify
the group theoretical background of the results obtained in the last section.
It is assumed below that the reader is well acquainted with the basic notions of
representation theory.\\

Classical 'fields' or 'wave functions' are spacetime-dependent functions or distributions serving for
the construction and description of a multiplicity of purely theoretical quantities or observables closely
related to the measurement process. The transformation property of fields transforming according to
\begin{equation}
\Psi'_{\lambda} (x') = D_{\lambda}^{\, \, \rho} (\Lambda)  \Psi_\rho (x) \, , \quad x'^\mu = \Lambda^\mu_{\, \, \nu} x^\nu \, ,
\label{recipe_inv}
\end{equation} 
where $D(\Lambda)$ is a $n \times n$-representation matrix which is associated with a not necessarily
irreducible (ray) representation of the proper orthochronous Lorentz group, will be referred to as
{\emph{manifestly Lorentz invariant}} in the following, if the defining (ray) representation property
\begin{equation}
D_\lambda^{\, \, \rho} (\Lambda_2 \Lambda_1) = \pm D_\lambda^{\, \, \alpha} (\Lambda_2) D_{\alpha}^{\, \, \rho} (\Lambda_1)
\end{equation}
holds. The sign appearing above which distinguishes ordinary representations from ray representations by allowing for a phase
in the homomorphism property may appear in in the case of the so-called spinor ray representations (of the proper
orthochronous Lorentz group $SO^+(1,3)$) defined below, a result which will
not be motivated any further in this paper \cite{Sexl}.\\

A well-known result from the representation theory of the rotation group states that all finite-dimensional irreducible
ray representations $\vartheta_{j}$ of $SO(3)$ (or all finite-dimensional {\emph{irreps}} of the corresponding universal covering group $SU(2)$)
can be labelled by $j=(n-1)/2$, where $n \in \mathds{N}$ is the dimension
of $\vartheta_j$, since all irreducible (ray) representations of a given dimension $n$ are unique {\emph{up to equivalence}}.
Furthermore, the tensor (or {\emph{Kronecker}, or {\emph{direct}) product of two such representations decays into a direct sum according to the
Clebsch-Gordan decomposition
\begin{equation}
\vartheta_{j_1} \otimes \vartheta_{j_2} = \bigoplus \limits_{j=|j_1-j_2|}^{j_1+j_2} \vartheta_j \, , \quad \mbox{or} \quad
j_1 \otimes j_2 = |j_1 - j_2 | \oplus |j_1-j_2+1| \oplus \ldots \oplus |j_1 + j_2| \, , \label{CGD}
\end{equation}
i.e., we denote representations up to equivalence by the symbol $\vartheta_j$ or directly by the 'quantum number' $j$ which
classifies the representation.
E.g., the tensor product of two spin-$\frac{1}{2}$-representations contains a spin-$0$ and a spin-$1$-representation
\begin{equation}
\vartheta_{\frac{1}{2}} \otimes \vartheta_{\frac{1}{2}} = \vartheta_0 \oplus \vartheta_1 \ , \quad \mbox{or} \quad
\frac{1}{2} \otimes \frac{1}{2} = 0 \oplus 1 \, .
\end{equation}
A $SU(2)$-representation $\vartheta_{j}$ and its complex conjugate representation $\vartheta_j^*$ are equivalent.\\

Going beyond the rotation group, one finds that all existing finite-dimensional (ray) irreps $\vartheta_{j,j'}$ of the group
($SO^+(1,3)$) $SL(2, \mathds{C})$ can be labelled by two indices $j,j' = 0,\frac{1}{2},1,\frac{3}{2}, \ldots$,
and the Clebsch-Gordan decomposition (\ref{CGD}) generalizes to
\begin{equation}
\vartheta_{j_1,j'_1} \otimes \vartheta_{j_2,j'_2} =  \bigoplus \limits_{J=|j_1-j_2|}^{j_1+j_2}
 \bigoplus \limits_{J'=|j'_1-j'_2|}^{j'_1+j'_2} \vartheta_{J,J'} \, . \label{LCGD}
\end{equation}
The $\vartheta_{j,j'}$ can be constructed as tensor products according to
\begin{equation}
\vartheta_{j,j'} = \vartheta_{j,0} \otimes \vartheta_{0,j'} = \vartheta_{j,0} \otimes \vartheta_{j',0}^* \, ,
\end{equation}
and consequently all $\vartheta_{j,j'}$ can be generated inductively from the fundamental representations $\vartheta_{0,\frac{1}{2}}$ and
$\vartheta_{\frac{1}{2},0}$; for this reason they are called fundamental.
Restricting the $SL(2, \mathds{C})$-representations $\vartheta_{j,0}$ and $\vartheta_{0,j}$ to the subgroup
$SU(2)$ leads to the $SU(2)$-representations $\vartheta_j$.
The analogy of the decomposition (\ref{LCGD}) with the $SU(2)$ case is rooted in the fact that the complex six-dimensional
Lie algebra of the complex Lorentz group $SO(4,\mathds{C})$ is the direct sum of two $SO(3,\mathds{C})$ Lie subalgebras,
which itself originates from the algebra of rotation and boost generators in the real case.
The complex dimension $n=\mbox{dim}_\mathds{C} (\vartheta_{j,j'})$ is given by $n=(2j+1)(2j'+1)$.
One should note that interchanging the indices $j$ and $j'$ according to
\begin{equation}
\vartheta_{j,j'} = \vartheta_{j',j}^*
\end{equation}
relates complex conjugate representations which are not equivalent for $j \neq j'$. The representations $\vartheta_{j,j}$ are real,
i.e. they can be represented by real $(2j+1)^2 \times (2j+1)^2$-matrices.
Only the trivial representation
$\vartheta_{0,0}$ of the Lorentz group is unitary. All other unitary irreps of the Lorentz group are infinite-dimensional
and are commonly constructed by the help of wave function spaces.\\

The following fields, transforming according to the lowest-dimensional, not necessarily irreducible (ray) representations $\vartheta_{j,j'}$
of the proper orthochronous Lorentz group, play the most important r\^oles in relativistic (quantum) field theory in
$3+1$-dimensional Minkowski spacetime:
\begin{itemize}
\item $(j,j')=(0,0)$: Real or complex scalar field $\varphi(x)$.
\begin{equation}
\varphi' (x') = D_{0,0}(\Lambda) \varphi(x) = \varphi(x) = \varphi(\Lambda^{-1} x') \, .
\end{equation}
\item $(j,j')=(\frac{1}{2},0)$: Complex two-component right-chiral spinor field $\psi_R (x)$.
\begin{equation}
\psi'_{R,\alpha} (x') = A_\alpha^{\, \, \beta} (\Lambda) \psi_{R,\beta} (x) 
=A_\alpha^{\, \, \beta} (\Lambda) \psi_{R,\beta} (\Lambda^{-1} x') \, , \quad A(\Lambda) = D_{\frac{1}{2},0} (\Lambda) \, .
\end{equation}
\item $(j,j')=(0, \frac{1}{2})$: Complex two-component left-chiral spinor field $\psi_L (x)$.
\begin{equation}
\psi'_{L,{\overline{\alpha}}} (x') = A_{\overline{\alpha}}^{* {\overline{\beta}}} (\Lambda) \psi_{L,{\overline{\beta}}} (x) 
=A_{\overline{\alpha}}^{* {\overline{\beta}}}
(\Lambda) \psi_{L,{\overline{\beta}}} (\Lambda^{-1} x') \, , \quad A^*(\Lambda) = D_{0,\frac{1}{2}} (\Lambda) \, .
\end{equation}
As a reminiscence to the literature using dotted and undotted spinor indices according to varying conventions,
two types of spinor indices were used above to distinguish between the two fundamental $SL(2, \mathds{C})$-representations.

\item $(j,j')=(\frac{1}{2},\frac{1}{2})$: Real or complex vector field $V_{\alpha \bar{\beta}}(x) / V^\mu(x)$.\\
One has $\vartheta_{\frac{1}{2}, \frac{1}{2}} = \vartheta_{\frac{1}{2},0} \otimes \vartheta_{0, \frac{1}{2}}$
or $(\frac{1}{2},\frac{1}{2}) = (\frac{1}{2},0) \otimes (0, \frac{1}{2})$, and
the transformation law following for $V_{\alpha \bar{\beta}}(x)$ under the direct product of the
representations $\vartheta_{\frac{1}{2},0}$ and
$\vartheta_{0,\frac{1}{2}}=\vartheta^*_{\frac{1}{2},0}$
\begin{equation}
V'_{\alpha \overline{\beta}} (x') = A_{\alpha}^{\, \, \gamma} A^{*  \, {\overline{\delta}}}_{{\overline{\beta}}}
V_{\gamma {\overline{\delta}}} ( \Lambda^{-1} x')
\end{equation}
can be cast into an interesting form by using the generalized Pauli matrices as a basis of the complex
vector space of the $2 \times 2$-matrices $Mat(2, \mathds{C})$ in order to define the four fields
$V^0$, $V^1$, $V^2$, and $V^3$ according to
\begin{equation}
V_{\gamma {\overline{\delta}}} (x) = \sum \limits_{\mu=0}^{3} V^\mu (x) ({\sigma_\mu})_{\gamma {\overline{\delta}}}
=V^\mu (x) ({\sigma_\mu})_{\gamma {\overline{\delta}}} \, \sim
\left(\begin{array}{cc}
V^0+V^3 & V^1-i V^2  \\
V^1 + i V^2 & V^0-V^3 \\
\end{array} \right) (x) \, ,
\end{equation}
leading to
\begin{displaymath}
V'^\mu ({\sigma_\mu})_{\alpha {\overline{\beta}}} (x')=
V'_{\alpha \overline{\beta}} (x') = A_{\alpha}^{\, \, \gamma} A^{*  \, {\overline{\delta}}}_{{\overline{\beta}}}
V_{\gamma {\overline{\delta}}} ( \Lambda^{-1} x')
\end{displaymath}
\begin{equation}
= \underbrace{(A V(x) A^+)_{\alpha \overline{\beta}}}_{
\left(\begin{array}{cc}
V'^0+V'^3 & V'^1-i V'^2  \\
V'^1 + i V'^2 & V'^0-V'^3 \\
\end{array} \right) (x') }
\overset{!}{=} (\Lambda^\mu_{\, \, \nu} V^\nu (x)) (\sigma_\mu)_{\alpha {\overline{\beta}}} \, .
\end{equation}
Indeed, $V^\mu$ is a {\emph{vector field}} and transforms like the spacetime coordinates.
The $V^\mu$ are not necessarily complex, as one knows from the relativistic four-potential $A^\mu$ in electrodynamics
or the (massive) classical Proca field $Z^\mu$ used to describe the classical $Z$-boson.
In the complex case, the vector field may be used to describe charged fields $W^{* \mu} \neq W^\mu$
and associated particles like the $W$-bosons.

\item $(j,j')=(1,0)$ oder $(0,1)$: Complex Riemann-Silberstein vector fields $\vec{F} (x) = \vec{E} (x) + i \vec{B} (x)$
or $\vec{F} (x)^* = \vec{E} (x) - i \vec{B} (x)$.\\
The direct sum of the representations $(1,0) \oplus (0,1)$ can be used to construct a six-dimensional real representation
of the Lorentz group which is linked to the Lorentz transformation properties of the electric and magnetic field
 $\vec{E} (x)$ and  $\vec{B} (x)$, respectively.

\item $(\frac{1}{2},0) \oplus (0, \frac{1}{2})$: Dirac spinors $\Psi(x)$.\\
Dirac spinors are used to describe the Standard Model
spin-$\frac{1}{2}$ particles, i.e. leptons and quarks. The representation $(\frac{1}{2},0) \oplus (0, \frac{1}{2})$
can be restricted to four real dimensions and leads to the concept of four-component Majorana fields. This observation
is one of the main subjects of this paper and will be elucidated below in further detail.
\end{itemize}

\subsection{Real (ray) representations of the proper ortho\-chronous Lorentz group}
Complex half-integer representations $\vartheta_{j,j'}$ with $j+j'=\frac{1}{2}, \frac{3}{2}, \frac{5}{2}, \ldots$ are called
\emph{spinor} ray representations of the proper orthochronous Lorentz group, integer representations
$\vartheta_{j,j'}$ with $j+j' \in \mathds{N}$ are {\emph{tensor}} representations. Spinor representations
are faithful representations of the $SL(2, \mathds{C})$.
The reduction formula (\ref{LCGD}) explicitly holds in the case of the {\emph{complex}} representation theory
of the groups $SL(2, \mathds{C})$ and $SO^+(1,3)$.
However, also {\emph{real}} irreducible representations play a crucial r\^ole in quantum mechanics in connection
with the description of neutral fields like, e.g., the Higgs field in the Standard Model, the real antisymmetric field strength tensor
in electrodynamics or the gravitational field.
Of course, these classical entities lead to states in corresponding (Fock-) Hilbert spaces after second quantization, and these states
can be superposed according to the manifest complex structure of quantum mechanics.\\

The {\emph{real irreducible}} $SL(2, \mathds{C})$-representations can be classified into two types \cite{Gelfand}:
\begin{itemize}
\item Type 1:  $\vartheta^\mathds{R}_{j}$ ($j=0,\frac{1}{2},1,\frac{3}{2}, \ldots$) is obtained from
restricting a complex representation $\vartheta_{j,j}$
acting on $\mathds{C}^{(2j+1)^2}_\mathds{C}$ to a real subspace which is isomorphic to
$\mathds{R}^{(2j+1)^2}_\mathds{R}$. A more suggestive notation used below for such representations obtained
from the complex irreps $(j,j)$ is $(j,j)_\mathds{R}$.
\item Type 2:  $\vartheta^\mathds{R}_{j,k} \simeq \vartheta^\mathds{R}_{k,j}$  with $j \neq k$ is obtained from
restricting the direct sum $\vartheta_{j,k} \oplus \vartheta_{k,j}$ of an $SL(2, \mathds{C})$ irrep and its complex conjugate
to the real subspace $\mathds{R}^{2(2j+1)(2k+1)}_\mathds{R} \subset \mathds{C}^{2(2j+1)(2k+1)}_\mathds{C}$.
From a 'complex point of view', such representations are reducible, but they are not reducible in the real sense.
These representations shall be denoted below by $(j,k)_\mathds{R} \simeq (k,j)_\mathds{R}$.
Having projected out such a real representation from $\vartheta_{j,k} \oplus \vartheta_{k,j}$, there remains a second equivalent
real representation with, of course, the same dimension; the total dimension of both real representations
is then $4(2j+1)(2k+1)=\dim_\mathds{R} \mathds{C}^{2(2j+1)(2k+1)}$.
\end{itemize}

Since the second type is directly linked to the group theory of Dirac and Majorana fields, this case shall be investigated in
the following pedestrian way. The real irreducible representations contained in the complex reducible
$SL(2,\mathds{C})$-representation $\vartheta_{j,k} \oplus \vartheta_{k,j}$ can be isolated by the following explicit calculations.
Let $R$ and $I$ denote the real and the imaginary part of the
$n \times n$-representation matrix $D(A)=R(A)+iI(A)$ with $A \in SL(2, \mathds{C})$ and
$n=(2j+1)(2k+1)$, corresponding to a given representation $(j,k)$.
Then, a $2n \times 2n$-representation matrix $\tilde{D}$ of the direct sum $(j,k) \oplus (k,j) = (j,k) \oplus (j,k)^*$
can be written as
\begin{equation}
\tilde{D}=
\left(\begin{array}{cc}
R + i I &  0\\
0 & R-iI 
\end{array}\right) \, .
\end{equation}
As a complex representation matrix, $\tilde{D}$ acts on complex $2n$-component column vectors in
$\mathds{C}^{2n}_\mathds{C}$. However, we now focus on the {\emph{real}} $2n$-dimensional
subspace spanned by vectors which can be represented in the form
\begin{equation}
v=
\left(\begin{array}{c}
v_1 + i v_2\\ v_1-i v_2
\end{array}\right) \, , \quad v_{1,2} \in \mathds{R}^n \, .
\end{equation}
Such vectors are {\emph{real}} linear combinations of the $2n$ basis vectors
\begin{equation}
\left(\begin{array}{c}
1 \\ 0 \\ 0 \\ \vdots \\ 1\\ 0\\ 0\\ \vdots
\end{array}\right) \, , \quad 
\left(\begin{array}{c}
0 \\ 1 \\ 0 \\ \vdots \\ 0\\ 1\\ 0\\ \vdots
\end{array}\right) \, , \, \ldots \, , \quad
\left(\begin{array}{c}
0 \\ 0 \\ \vdots \\ 1 \\ 0\\ 0 \\ \vdots \\ 1
\end{array}\right) \, , \, \ldots \, , \quad
\left(\begin{array}{c}
i \\ 0 \\ 0 \\ \vdots \\ -i\\ 0\\ 0\\ \vdots
\end{array}\right) \, , \quad 
\left(\begin{array}{c}
0 \\ i \\ 0 \\ \vdots \\ 0\\ -i\\ 0\\ \vdots
\end{array}\right) \, , \, \ldots \, , \quad
\left(\begin{array}{c}
0 \\ 0 \\ \vdots \\ i \\ 0\\ 0 \\ \vdots \\ -i
\end{array}\right) \, .
\end{equation}
When $\tilde{D}$ acts as a linear operator on such a vector $v$, one obtains
\begin{equation}
\tilde{D} v =
\left(\begin{array}{cc}
R + i I &  0\\
0 & R-iI 
\end{array}\right)
\left(\begin{array}{c}
v_1 + i v_2\\ v_1-i v_2
\end{array}\right) =
\left(\begin{array}{c}
Rv_1 -  I v_2 + i(Iv_1 +Rv_2) \\ 
Rv_1 -  I v_2 - i(Iv_1 +Rv_2)
\end{array}\right) \, .
\end{equation}
This result can be immediately translated into a real representation defined by real
matrices $\hat{D}$
\begin{equation}
\left(\begin{array}{c}
v_1 \\ v_2
\end{array}\right) \mapsto 
\hat{D} \left(\begin{array}{c}
v_1 \\ v_2
\end{array}\right)
=
\left(\begin{array}{cc}
R &  -I\\
I & R 
\end{array}\right)
\left(\begin{array}{c}
v_1 \\  v_2
\end{array}\right)
=
\left(\begin{array}{c}
R v_1- I v_2 \\ I v_1 + R v_2
\end{array}\right) \, ,
\end{equation}
which display the multiplicative (homomorphism) representation property of their complex counterparts:
\begin{displaymath}
\tilde{D}_{2,1} =
\tilde{D}_2 \tilde{D}_1 =
\left(\begin{array}{cc}
R_2 + i I_2 &  0\\
0 & R_2-iI_2 
\end{array}\right)
\left(\begin{array}{cc}
R_1 + i I_1 &  0\\
0 & R_1-iI_1 
\end{array}\right) 
\end{displaymath}
\begin{equation}
=
\left(\begin{array}{cc}
R_2 R_1 -I_2 I_1 + i (R_2 I_1 +I_2 R_1)  &  0\\
0 & R_2 R_1 - I_2 I_1 -i (R_2 I_1 +I_2 R_1) 
\end{array}\right)
\end{equation}
becomes in the real case
\begin{equation}
\hat{D}_{2,1} =
\hat{D}_2 \hat{D}_1 =
\left(\begin{array}{cc}
R_2 &  -I_2\\
I_2 & R_2 
\end{array}\right)
\left(\begin{array}{cc}
R_1 &  -I_1\\
I_1 & R_1 
\end{array}\right)
=
\left(\begin{array}{cc}
R_2 R_1 - I_2 I_1 &  -R_2 I_1 - I_2 R_1\\
R_2 I_1 + I_2 R_1 & R_2 R_1 - I_2 I_1   
\end{array}\right)  \, .
\end{equation}

E.g., considering the Kronecker product of two vector representations, one has
in the complex case from decomposition (\ref{LCGD}) in compact notation
\begin{equation}
(\frac{1}{2}, \frac{1}{2}) \otimes (\frac{1}{2}, \frac{1}{2}) = (0,0) \oplus (1,0) \oplus (0,1) \oplus (1,1) \, .
\end{equation}
Restricting this complex result to the real content leads to a sum of two type 1 real representations
($(0,0)$ and $(1,1)$) and a type 2 real representation
($(1,0) \oplus (0,1)$)
\begin{equation}
((\frac{1}{2}, \frac{1}{2}) \otimes (\frac{1}{2}, \frac{1}{2}))_\mathds{R} =
(0,0)_\mathds{R} \oplus (1,0)_\mathds{R} \oplus (1,1)_\mathds{R} \, .
\end{equation}
For the direct product of the real representations one has therefore
\begin{equation}
\underbrace{\vartheta_{\frac{1}{2}}^\mathds{R} \otimes \vartheta_{\frac{1}{2}}^\mathds{R}}_{\mbox{dim}_\mathds{R}=16} = 
\underbrace{\vartheta_{0}^\mathds{R}}_{\mbox{dim}_\mathds{R} = 1} \oplus
\underbrace{\vartheta_{1,0}^\mathds{R}}_{\mbox{dim}_\mathds{R} = 6} \oplus 
\underbrace{\vartheta_{1}^\mathds{R}}_{\mbox{dim}_\mathds{R}=9} \, .         \label{real_dec}
\end{equation}

This decomposition is best illustrated by investigating a real second rank tensor field $T^{\mu \nu}(x)$,
which transforms according to the tensor product of the real representation of the proper orthochronous Lorentz group
$SO^+(1,3)$ by itself and with itself according to
\begin{equation}
T'^{\mu \nu} (x') = \Lambda^\mu_{\, \, \alpha} \Lambda^\nu_{\, \, \beta} T^{\alpha \beta} (x) \, ,
\label{double_trafo}
\end{equation}
i.e., according to $(\frac{1}{2},\frac{1}{2})_\mathds{R} \otimes (\frac{1}{2},\frac{1}{2})_\mathds{R}$.
$T^{\mu \nu}(x)$ can be decomposed in a unique manner into a spacetime-dependent part which is proprtional to
the inverse metric tensor
$g^{\mu \nu} = diag(1,-1,-1,-1)$, an antisymmetric tensor $F^{\mu \nu}(x) = -F^{\nu \mu}(x)$ and
a traceless symmetric tensor $H^{\mu \nu} (x) = H^{\nu \mu} (x)$ with
$H^\mu_{\, \, \mu} (x) = g_{\mu \nu} H^{\mu \nu}(x)=0$, as follows
\begin{equation}
T^{\mu \nu} (x) = \varphi(x) g^{\mu \nu} + F^{\mu \nu} (x) + H^{\mu \nu} (x) \, ,
\end{equation}
where (note that $g^\mu_{\, \, \mu} = \delta^\mu_{\, \, \mu} = 4$)
\begin{equation}
\varphi(x) = \frac{1}{4}  T^{\mu}_{\, \,  \mu} (x) \, , \quad F^{\mu \nu} (x) = \frac{1}{2} ( T^{\mu \nu} (x) - T^{\nu \mu} (x)) \, ,
\quad H^{\mu \nu} (x) = \frac{1}{2} (T^{\mu \nu} (x) + T^{\nu \mu} (x)) - \varphi(x) g^{\mu \nu} \, .
\end{equation}
The antisymmetric tensor field $F^{\mu \nu}$ is given by $6$ real spacetime-dependent field components transforming under
the $(1,0)_\mathds{R}$-representation;
the traceless symmetric tensor field $H^{\mu \nu}$ contains $9$ independent real field components ($\rightarrow (1,1)_\mathds{R}$),
and the component in $T^{\mu \nu}$ which is
proportional to the inverse metric tensor is related to the real scalar field $\varphi(x)$ ($ \rightarrow (0,0)_\mathds{R}$),
everything in accordance with the decomposition displayed in (\ref{real_dec}).\\

In matrix notation, the transformation (\ref{double_trafo}) can be expressed by $T'=\Lambda T \Lambda^T$,
and the trace $T^\mu_{\, \, \mu}$ becomes $\mbox{tr} \, (T g)$. Obviously, due to the defining property
$\Lambda^T g \Lambda =g$ of the matrices in $O(1,3)$
\begin{equation}
\mbox{tr} \, (T'g) = \mbox{tr} ( \Lambda T \Lambda^T g) = \mbox{tr} (T \Lambda^T g \Lambda) = \mbox{tr} \, (Tg)
\end{equation}
holds, i.e., the trace of a second rank tensor is a Lorentz invariant scalar.\\

Having all this group theoretical tools in our backpack, the obervations elaborated in the last section by explicit calculations
now receive a simple explanation. Coupling a left- and a right-handed chiral (Weyl) spinor field by mass terms
as performed in eqns. (\ref{L}) and (\ref{R}) imposes an additional dynamics on the total field of four complex or eight real field
components. In the Dirac case, the mass spectrum is degenerate and the structure of the field equations remains complex
such that the field components transform according to the reducible complex representation
$(\frac{1}{2},0) \oplus (0, \frac{1}{2})$; complex linear superpositions of solutions of the field equations are still solutions.
In the Majorana case, the representation splits up into two equivalent real four-dimensional representations
$(\frac{1}{2},0)_\mathds{R}$ of type 2,
with representation spaces which are invariant under the proper orthochronous Lorentz group,
containing two real four-component Majorana fields with independent dynamics imposed by the equations
of motion and independent Majorana masses.

\section{Conclusions}
In this paper, a comprehensive derivation and concise discussion of the free field wave equations governing
the dynamics of the fundamental two-component and four-component spin-$\frac{1}{2}$ matter fields
in Minkowski spacetime is presented.
The discussion is solely based on first principles like Lorentz symmetry, locality, causality, and unitarity
which result in the hyperbolic differential equations describing Weyl, Dirac or
Majorana fields. Coupling a fundamental two-component left-chiral field with a two-component right-chiral field
in the most general non-trivial way leads to Dirac fields or Majorana fields and an emergent Majorana phase.
A pure matrix-based formalism is used, avoiding an explicit van der Waerden notation \cite{Waerden}
with dotted and undotted spinor indices sometimes confusing researchers from different fields which are more
familiar with a notation inspired from linear algebra.

\end{document}